# Synthesis, crystal structure, polymorphism and microscopic luminescence properties of anthracene derivative compounds

Anna Moliterni[a]*, Davide Altamura[a], Rocco Lassandro[a], Vincent Olieric[b], Gianmarco Ferri[c], Francesco Cardarelli[c], Andrea Camposeo[d], Dario Pisignano[de], John E. Anthony[f] and Cinzia Giannini[a]*

[a]Istituto di Cristallografia, CNR, Via Amendola, 122/O, Bari, 70126, Italy

[b] Paul Scherrer Institute, Forschungstrasse 111, Villigen-PSI, 5232, Switzerland

[c]NEST, Scuola Normale Superiore, Piazza San Silvestro 12, Pisa, I-56127, Italy

[d]NEST, Istituto Nanoscienze, CNR, Piazza San Silvestro 12, Pisa, I-56127, Italy

[e]Dipartimento di Fisica 'E. Fermi', University of Pisa, Pisa, I-56127, Italy

[f]Center for Applied Energy Research, University of Kentucky, Research Park Drive, Lexington, KY, 2582, USA

Correspondence email: annagrazia.moliterni@ic.cnr.it; cinzia.giannini@ic.cnr.it

**Abstract**   Anthracene derivative compounds are currently investigated because of their unique physical properties (*e.g*., bright luminescence and emission tunability), which make them ideal candidates for advanced optoelectronic devices. Intermolecular interactions are the basis of the tunability of the optical and electronic properties of these compounds, whose prediction and exploitation benefit from the knowledge of the crystal structure and the packing architecture. Polymorphism can occur due to the weak intermolecular interactions, asking for detailed structural analysis clarifying the origin of observed material property modifications. Here, two silylethyne-substituted anthracene compounds are characterized by single-crystal synchrotron X-ray diffraction, identifying a new polymorph. Additionally, laser confocal microscopy and fluorescence lifetime imaging microscopy confirm the results obtained by the X-ray diffraction characterization, *i.e*., shifting the substituents towards the external benzene rings of the anthracene unit favours $\pi$-$\pi$ interactions, impacting on both the morphology and the microscopic optical properties of the crystals. The compounds with more isolated anthracene units feature shorter lifetime and emission spectra more similar to those ones of isolated molecules. The crystallographic study, supported by the optical investigation, sheds light on the influence of non-covalent interactions on the crystal packing and luminescence properties of anthracene derivatives, providing a further step towards their efficient use as building blocks in active components of light sources and photonic networks.

**1. Introduction**

Organic semiconductors have known a growing interest during the last few decades due to their exploitation as active layers of a new generation of optoelectronic devices, such as organic light-emitting diodes (OLEDs; Yersin, 2008; Liu *et al*., 2013), organic solar cellar (OSCs; Palilis, *et al*., 2008; Dou *et al*., 2013; Ostroverkhova, 2016), organic field-effect transistors (OFETs; Ito *et al*., 2003; Allard *et al*., 2008; Wang *et al*, 2012; Mei *et al*., 2013). Understanding their optoelectronic properties and the correlations with mechanical and morphological properties opened the door to the development of mechanical flexible, easily produced and cheap components for photonics, electronics, and energy conversion (Griffith *et al*., 2010).

Among the organic semiconductors, acenes are aromatic hydrocarbons consisting of linearly fused benzene rings; the smallest compound of the acene family is anthracene that can be easily obtained from anthracene oil, *i.e*., the coal-tar fraction that distils at a temperature above 270 °C. These compounds are especially interesting in their crystalline forms, which allow intriguing effects to be observed, such as polariton lasing (Kena-Cohen *et al.*, 2010). Intermolecular interactions play a major role in organic crystalline materials, not only determining molecular packing and in turn the optical and electrical properties of the solid-state material, but also its macroscopic properties, such as crystal habit, directly affecting light polarization, confinement and guiding (Camposeo *et al*., 2019). For example, needle-shaped crystals can exhibit optical waveguide properties with low propagation losses (Camposeo *et al*., 2019). $\pi$-$\pi$ interactions favour an efficient channel for charge mobility (Anthony *et al*., 2002; Chen *et al*., 2006; Yao *et al*., 2018), and are therefore exploited in material design (da Silva Filho *et al*., 2005). It is also known that non-covalent self-assembly of acenes can be controlled by the insertion of suitable molecular substituents, by which crystal properties can be effectively tailored (Anthony, 2005, 2008). Therefore, engineering of the active material relies on the in-depth knowledge of crystal structure and non-covalent interactions. Here, we provide such a detailed characterization through single crystal synchrotron X-ray diffraction for two silylethyne-substituted anthracene compounds, *i.e*., 1,2,3,4-Tetrafluoro-5,8-bis(trimethylsilylethynyl)anthracene and 9,10-bis(triisopropylsilylethynyl)anthracene (F4 TMS ANT and TIPS ANT, respectively), previously studied by Camposeo *et al.* (2019), and Anthony and Parkin (2016), respectively. The structural characterization of F4 TMS ANT carried out by Camposeo *et al*. (2019) is confirmed and improved in terms of structure model accuracy, and a detailed description of the intermolecular interactions is provided as well. In addition, a new polymorph of TIPS ANT is identified (named TIPS ANT*p* in this work), revealing that different molecular packing can result from the same synthesis procedure (Landis *et al*., 2005). The occurrence of polymorphism in case of TIPS Anthracene compounds has been recently observed also by Bhattacharyya and Datta (2017). The position of the molecular substituents towards the external benzene rings is shown to influence the crystal morphology (*i.e.,* needle and plate shape in the case of F4 TMS ANT and TIPS ANT*p*, respectively) and the strengths of the $\pi$-$\pi$ interactions.

Laser confocal microscopy and fluorescence lifetime imaging microscopy show for the two compounds significant differences in the luminescence properties, along with the uniform emission throughout both the compounds. The lifetime measurements here performed evidence decay times of the photoluminescence (PL) an order of magnitude longer for F4 TMS ANT with respect to TIPS ANT*p*. The unlike PL features of the two compounds are in agreement with their dissimilar molecular arrangement suggested by the crystallographic study.

The accurate structural description here provided is a further step in view of tailoring crystal morphology and optical properties to achieve the sought compromise between molecular stabilization and optimal performance of the organic semiconductor (Gu *et al*., 2012) and/or its coupling to an optical network (Camposeo *et al*., 2019).

## 2. Experimental

### 2.1. Synthesis and crystallization

Synthesis of F4 TMS ANT has been described in (Camposeo *et al*., 2019); the synthesis method for obtaining TIPS ANT*p* has been reported by Landis and co-workers (Landis *et al*., 2005).

### 2.2. X-ray diffraction

Single-crystal X-ray diffraction data were collected at the beamline PXIII (X06DA-PXIII, http://www.psi.ch/sls/pxiii/) at the Swiss Light Source (SLS), Villigen, Switzerland, using a Parallel Robotics Inspired (PRIGo) multi-axis goniometer (Waltersperger *et al*., 2015) and a PILATUS 2M-F detector. Data collections were performed at room temperature (T = 296 K) on selected crystals of F4 TMS ANT and TIPS ANT*p*, mounted on litholoops (Molecular Dimensions). For each crystal, complete data were obtained by merging two 360º ω scans at χ=0° and χ =30° of PRIGo. In shutterless mode, a 360º data set was collected in 3 min (beam energy of 17 keV, λ= 0.72932 Å, focus size 90 × 50 μm$^2$, 0.25 sec of exposure time per frame, 0.5° scan angle).

Main data collection details are given in Table 1. Partial data sets were individually processed by *XDS* (Kabsch, 2010), a software organized in eight subroutines able to carry out the main data reduction steps; the corresponding XDS_ASCII.HKL reflection files were scaled and merged by the *XSCALE* subroutine (Kabsch, 2010). Structure solution was carried out by Direct Methods using *SIR2019* (Burla *et al*., 2015) and refined by *SHELXL2014/7* (Sheldrick, 2015). All non-hydrogen atoms were refined anisotropically. The carbon-bound H atoms were placed on geometrically calculated positions and refined using a riding-model approximation.

### 2.3. Laser Confocal Microscopy and Fluorescence Lifetime Imaging Microscopy

The microscale optical properties of F4 TMS ANT and TIPS ANT*p* were investigated by laser confocal microscopy (LCM). To this aim, spatially-resolved emission spectra were measured by an inverted microscope equipped with a laser confocal scanning head (FV1000, Olympus), by exciting samples with a continuous wave laser emitting at 405 nm through either a 10× objective (Olympus UPLSAPO) with 0.4 numerical aperture (NA) or a 60× objective (Olympus PLAPON) with NA=1.42. The excitation power was in the range 10-50 µW. The laser was focussed to a diffraction-limited spot onto the sample. The photoluminescence was collected by the same objective and measured by a photomultiplier (Olympus). The sampling speeds during the measurements of the fluorescence micrographs were in the interval 15-80 µs/µm. Typically, higher excitation power and sampling speed were used for F4 TMS ANT samples. In order to measure the spatially-resolved PL spectra, confocal micrographs were collected at various emission wavelengths with a spectral bandwidth of 2 nm. A polarization analyser positioned on the optical path of the emission was used for measuring polarized PL spectra. To this aim, samples were excited by a linearly polarized laser, with polarization direction parallel to the long axis of the TIPS ANT*p* platelets. The PL lifetime for the crystalline samples was investigated by confocal fluorescence lifetime imaging microscopy (FLIM). This analysis was carried out by an inverted microscope with confocal head (TCS SP5, Leica Microsystem) and either a 40× (NA = 1.25) or a 100× objective (Fluotar, NA = 1.3). Samples were optically excited by a 405 nm pulsed diode laser (Picoquant, Sepia Multichannel picosecond diode laser, maximum average power 30 µW) operating at 40 MHz, whereas the fluorescence intensity was measured by a photomultiplier tube interfaced with a Time Correlated Single Photon Counting setup (PicoHarp 300, PicoQuant, Berlin). The detection rate was kept in the interval $10^5$-$10^6$ counts/s, while lifetime time signals were time-integrated until reaching an average value of the order of $10^2$ counts in each area of the scanned micrographs. Spectral filters with variable bandwidths were exploited in order to measure the PL lifetime in different spectral intervals. The measured temporal decay profiles of the PL were fitted to exponential functions convoluted with the instrumental response function, that is assumed to be Gaussian with full width at half maximum of 280 ps.

For confocal analysis the crystalline samples were placed on top of a glass coverslip (thickness 150 µm), while for macroscopic optical characterization the crystalline samples were placed on the surface of a 1×1 cm² quartz substrates (thickness 1 mm). Crystalline samples with low inhomogeneities as visible by bright and dark field optical microscopy were selected for the measurements. Absorption spectra were measured by using a UV-visible spectrophotometer (Lambda 950, Perkin Elmer). The samples were mounted on a sample holder for solid state specimens that is made by a metallic plate with a central clear part and two clamps which block two edges of the quartz substrate, while leaving the central part of the substrate free for optical access. The incident optical beam was properly masked in order to have a spot size matching the area of the crystalline sample.

**2.4. Structure solution and refinement**

Both F4 TMS ANT and TIPS ANT*p* were solved and refined by single-crystal synchrotron X-ray diffraction data. According to the literature, TIPS ANT crystallized in the centrosymmetric space group *Pbca* and the related crystallographic data were deposited at the Cambridge Crystallographic Data Centre (CCDC), with deposition number CCDC 962668 (Anthony and Parkin, 2016); in case of TIPS ANT, data collection was carried out at a safe temperature of T= 250 K because the authors observed that a destructive phase transition occurred for crystals cooled to 240 K. In the present work, in order to investigate the occurrence of phase transitions caused by cooling, for TIPS ANT*p* two diffraction experiments were carried out, at room temperature and at 250 K, respectively. The analysis of the corresponding sets of diffraction data revealed no changes in the crystal structure. Consequently, the results here presented concern only room temperature measurements.

The structure characterization presented in this work in case of F4 TMS ANT confirmed the crystal structure results described by Camposeo *et al.* (2019), [whose corresponding CIF file was deposited at the Cambridge Crystallographic Data Centre (CCDC) with deposition number CCDC 1838578 and for the sake of completeness provided also as supplementary material (*i.e.*, file 1838578.cif)] while for TIPS ANT*p* enabled to identify a new polymorph of TIPS ANT, that crystallized in the centrosymmetric space group *P*-1.

The crystal structure solution step was carried out by *SIR2019* (Burla *et al.*, 2015) that exploits the information on cell parameters, diffraction intensities and expected chemical formula to determine the space group and solve the structure by Direct Methods (Giacovazzo, 2014). Crystal structures were refined using full-matrix least-squares techniques by *SHELXL2014/7* (Sheldrick, 2015). Non-hydrogen atoms were refined anisotropically while hydrogen atoms were geometrically positioned and constrained to ride on their parent C atoms with the following bond lengths constraints: C−H=0.96 Å and C−H = 0.93 Å for methyl and aromatic H atoms, respectively. The isotropic *U* value constraint $U_{iso}(H)=k\ U_{eq}(C)$ was set, with $k$=1.5 and 1.2 for methyl and aromatic H atoms, respectively; a rotating group model was applied for methyl groups.

Main crystal data and details on data collection and structure refinement are summarized in Table 1 that, in case of F4 TMS ANT, provides also the results obtained by Camposeo *et al.* (2019) (see the corresponding CIF file) for the sake of comparison.

**3. Results and discussion**

F4 TMS ANT crystallized in the centrosymmetric space group $P2_1/c$, with one molecule in the asymmetric unit (see Figure 1), and TIPS ANT*p* crystallized in the centrosymmetric space group *P*-1, with half a molecule in the asymmetric unit (see Figure 2). In case of F4 TMS ANT the space group determination was automatically carried out by *SIR2019* by considering the Laue group compatible with the geometry of the unit cell and assigning a probability value to each related extinction symbol, taking into account a statistical analysis carried out on the experimental intensities; at the end of this step the most plausible space group was graphically selected. For both compounds, the structure solution process was automatically performed by *SIR2019*. Crystal structure refinement was carried out by *SHELXL2014/7* by applying full-matrix least-squares techniques. Non-hydrogen atoms were refined anisotropically while a riding-model approximation was applied in case of hydrogen atoms: H atoms were geometrically positioned at the bond distances C−H=0.96 Å and C−H=0.93 Å for methyl and aromatic H atoms, respectively and allowed to ride on their respective parent C atoms. In case of methyl group, a rotating group model was assumed and the torsion angle defining its orientation about the Si−C bond [in case of F4 TMS ANT] or the C−C bond [in case of TIPS ANT*p*], was refined. The isotropic $U$ value satisfied the following constraints: $U_{iso}(H)=k\ U_{eq}(C)$, with $k$=1.5 and 1.2 for methyl and aromatic H atoms, respectively.

Main crystallographic data are given in Table 1; additional tables, concerning refined fractional atomic coordinates and displacement parameters, bond distances and angles, and torsion angles, were provided in the Supplementary Information.

For both compounds no presence of classical hydrogen bonds was detected; the crystal packing suggested that the main intermolecular interactions were weak and, in case of F4 TMS ANT, consisted of $\pi$-$\pi$ interactions (Janiak, 2000; Meyer *et al*., 2003; Martinez and Iverson, 2012) and $C_{aryl}$−F⋯H−C interactions (Meyer *et al*., 2003); for TIPS ANT*p*, they were aromatic interactions, weaker than those involved in F4 TMS ANT, and C−H⋯$\pi$ interactions (Meyer *et al*., 2003; Nishio, 2004; Nishio *et al*., 2012).

In case of F4 TMS ANT, the refined crystal structure here described is similar to that obtained by Camposeo *et al*. (2019), as shown in Figure 3, providing the overlay of the two structure solutions, having an r.m.s. deviation of 0.04 Å (*SIR2019*; Burla *et al*., 2015). Our results, compared to the structure model by Camposeo *et al*. (2019), are characterized by a not negligible improvement in terms of C−C bond precision and agreement factors; the bond precision, in case of our results, is 0.0032 Å instead of 0.0061 Å, while the $R[F^2 > 2\sigma(F^2)]$, $wR(F^2)$ were 0.048 and 0.148 instead of 0.062 and 0.175, respectively (see Table 1).

For F4 TMS ANT (Figure 1) the anthracene core consists of three planar-like six-membered rings. The mean distance from the least-squares plane calculated for 18 atoms, including 14 C-atoms of anthracene group and the four F-atoms, was 0.0074 Å and the largest deviation was at C14 (0.023 Å) and at F4 (0.019 Å). A slight bending was observed for the lateral chains involving the alkyne group: the distances from the least-squares plane in case of C15, C16 and Si1 were 0.024, 0.047 and 0.117 Å, respectively, while in case of C20, C21 and Si2 were 0.074, 0.135 and 0.280 Å, respectively. As usual for anthracenyl units, the peripheral aromatic rings were distorted from the hexagonal geometry with shortening of some bond distances (Kovalski *et al*., 2018): the bond distances C5−C6, C7−C8, C1−C14, C12−C13 were 1.34, 1.346, 1.378 and 1.364 Å, respectively, while the rest of bond distances were between 1.389 and 1.445 Å. The mean value of the C−C bonds was 1.4098 Å (ring C1 C2 C11 C12 C13 C14), 1.4042 Å (ring C2 C3 C4 C9 C10 C11) and 1.3938 Å (ring C4 C5 C6 C7 C8 C9). The crystal packing evidenced the presence of weak $C_{aryl}$−F···H−C interactions (Meyer *et al.*, 2003) and parallel offset π-π interactions (Janiak, 2000; Meyer *et al.*, 2003; Martinez and Iverson, 2012), as shown in Figures 4 and 5. These kinds of π-π arrangements are energetically more stable and favoured than the parallel face-centred stacked ones (Martinez and Iverson, 2012). The minimum distance between centroids of the aromatic rings was 3.698 Å (see Figure 5) and the minimum interplanar distance was 3.422 Å, in agreement with the typical values of the interplanar distances for π-π interactions, belonging to the range 3.3-3.8 Å (Janiak, 2000). The parallel offset π-π interactions were responsible for stacking arrangements along *a* (see Figure 4). $C_{aryl}$−F···H−C interactions contributed to stabilize the crystal structure (Meyer *et al.*, 2003) and the values of the distance F·· C belonged to the range observed for this kind of interactions (*i.e.*, 3.3-3.6 Å, see Meyer *et al.*, 2003).

In case of TIPS ANT*p*, the mean distance from the least-squares plane, calculated for the C-atoms of anthracene group, was 0.0041 Å and the largest deviation was at C3 (0.006 Å) and at C4 (0.006 Å). The bending observed for the lateral chain involving the alkyne group was larger than in case of F4 TMS ANT, probably due to the larger number of methyl groups: the distances from the least-squares plane in case of C6, C1 and Si1 were 0.048, 0.136 and 0.409 Å, respectively. Also for TIPS ANT*p* the peripheral aromatic rings were distorted from the hexagonal geometry: the bond distances C10−C12 and C9−C11 were 1.358 and 1.356 Å, respectively, while the rest of bond distances were between 1.411 and 1.430 Å. The mean value of the C-C bonds in the three aromatic rings shown in Figure 2 is 1.4005 Å (ring C4 C10 C12 C11i C9i C3i), 1.4173 Å (ring C3 C2 C4 C3i C2i C4i) and 1.4005 Å (ring C11 C9 C3 C4i C10i C12i). A view along *a* of the crystal packing is given in Figure 6. Differently from TIPS ANT [Anthony and Parkin (2016)], the crystal packing in TIPS ANT*p* (present work) did not reveal any edge-to-face interaction; the main intermolecular contacts were weak interactions between parallel aromatic rings and C−H···π interactions (Meyer *et al.*, 2003; Nishio, 2004; Nishio *et al.*, 2012). The last ones, as observed by Nishio (2004), are entropically favoured and contribute to stabilize the crystal structure. The parallel aromatic rings are characterized by a large offset (the shortest distance between centroids of parallel aromatic rings is 4.726 Å, see Figure 7, and the minimum interplanar distance is 2.471 Å); consequently, these interactions, in spite of the short interplanar distance, were weaker than the parallel-offset π-π interactions detected in F4 TMS ANT, and due to the large distance between centroids are not the typical π-π interactions (Janiak, 2000). If compared with F4 TMS ANT, the weaker π-π interactions in case of TIPS ANT*p* could lead to a reduction of the charge mobility, which should be confirmed by proper electrical characterization (this study is beyond the goal of this paper). The values of C-H···π distances shown in Figure 7 belonged to the distance range observed for this kind of interactions (*i.e.*, 3.3−4.1 Å, see Meyer *et al.*, 2003, Hattab *et al.*, 2010).

The main intermolecular interactions in case of F4 TMS ANT and TIPS ANT*p* crystal structures can be represented *via* Hirshfeld surfaces, using the *CrystalExplorer17* software (Turner *et al*., 2017). The Hirshfeld surface offers a useful tool for measuring the space occupied by a molecule in a crystal and summarizing information on all intermolecular interactions. In Figures S1 and S2 the Hirshfeld surface mapped over $d_{norm}$ is shown in case of F4 TMS ANT and TIPS ANT*p*, respectively. The conventions for the surface colours are the following ones: blue, white and red colours identify the interatomic contacts as longer, at van der Waals separations and short, respectively. In both the figures the blue colour is predominant. No red region is observed for TIPS ANT*p* while in case of F4 TMS ANT a very small and faint red region, indicated by red arrows in Figure S1a and zoomed in Figure S1b, corresponds to the $C_{aryl}$−F···H−C interactions.

The different arrangement and molecular interactions of the anthracene derivatives in the crystalline samples also affect the PL of the single crystals. Figure 8 compares the microscopic PL emission spectra of individual single crystals of F4 TMS ANT and TIPS ANT*p*, as measured by laser confocal microscopy. TIPS ANT*p* crystals feature brighter emission, typically requiring lower laser power for optical excitation (by about a factor 5) and sampling speeds (by about a factor 5) compared to F4 TMS ANT in order to obtain micrographs with comparable PL intensity per pixel. The broad and featureless PL spectrum of F4 TMS ANT is peaked at about 556 nm, with a full width at half maximum of 104 nm (Figure 8a); this maximum is red-shifted with respect to the emission maxima observed in case of TIPS ANT*p* (see Figure 8b), a feature due to π-π interactions between anthracene moieties, in accordance with literature [see, *f.e*., Teka *et al*. (2015)]. Spatially-resolved PL measurements were performed on single crystals of both samples (Figures 8c,d) with length of about 150 μm. F4 TMS ANT showed highly uniform emission spectral features as well as intensity along the length of the needle-like system (Figure 8c). Indeed, the variation ($\Delta\lambda_p$) of the PL peak wavelength ($\lambda_p$) along the length of the needle of F4 TMS ANT is <3 nm ($\Delta\lambda_p/\lambda_p$=0.5%), and a similar spatial stability is found for the PL full width at half maximum (measured values are in the interval 102-104 nm). The PL spectrum of a single platelet crystal of TIPS ANT*p* is, instead, much more structured, with peaks at 454 nm, 474 nm, 500 nm, 508 nm and 538 nm (Figure 8b). Some of these PL peaks (the ones at 474 nm and 508 nm) are close to those of the PL spectrum of molecules of TIPS ANT*p* in solution (see Figure S3 in the Supporting Information, where PL peaks at 446 nm, 475 nm and 507 nm can be identified for the molecule in solution) (del Valle *et al*., 2002). In addition, polarized PL spectra shown in Figure S4 in the Supporting Information highlight a variation of the shape of the emission spectrum of TIPS ANT*p* upon changing the polarization of collected light. This analysis unveils the presence of a peak, in the high energy tail of the PL spectrum, that is polarized in a direction parallel to the short axis of the crystal face, and a dependence of the intensity ratio of the peaks at 500 nm and 508 nm on polarization. The structured shape of the spectrum and the polarization dependence are indicative of the presence of different emissive species. Here we point out that the confocal microscopy measurements allow some spectral features of the spectrum below 475 nm to be unveiled for crystalline samples of TIPS ANT*p*, which were previously masked by self-absorption (Camposeo *et al*. 2019). This is highlighted in Figure S5a where the emission spectra of TIPS ANT*p*, measured by vertically shifting the high-numerical objective (NA=1.42) along the crystal thickness (*z* axis in Figure S5), are compared. This analysis highlights the increasing contribution of self-absorption as the excitation focal spot is shifted into the crystal, resulting in a decrease of the intensity of the high energy transitions with respects to the low energy one (Figure S5b). The larger self-absorption effect in case of TIPS ANT*p* is also favoured by the strong overlap between the absorption and PL spectra, more prominent compared to F4 TMS ANT (see Figure S6). Similarly to F4 TMS ANT, the PL spectrum does not feature significant variations within individual platelet crystal, as obtained by spatially-resolved photoluminescence (Figure 8d). The different shape of the crystalline samples (needle *vs* platelet) was demonstrated to determine a different light transport behaviour (Camposeo *et al*. 2019), namely a more efficient self-waveguiding of the emitted light in needles compared to platelets.

The PL lifetime of F4 TMS ANT and TIPS ANT*p* are shown in Figure 9a and 9b, respectively. While a long lifetime in the range 50-80 ns is estimated for F4 TMS ANT, the PL lifetime of TIPS ANT*p* is about an order of magnitude shorter, being in the interval 1-4 ns (compare insets of Figure 9a,b). The longer PL lifetime in case of F4 TMS ANT could be due to $\pi$-$\pi$ interactions and the columnar stacking of chromophores. Interestingly, also the spectral dependence of the lifetime is different for the two compounds: the temporal decay of the PL of F4 TMS ANT is constant throughout the spectral range of the emission, whereas for TIPS ANT*p* we observed an increase of the lifetime by red-shifting the PL wavelengths (Figures 9a,b). More specifically, the lifetime for the high energy components of TIPS ANT*p* (in the interval 450-475 nm) is about 0.8 ns, and it increases to about 4 ns for the low energy ones (525-600 nm). For the sake of comparison, we recall that solutions of TIPS ANT molecules at low concentrations have been reported to show a wavelength-independent lifetime of about 6.6 ns (Pun *et al.* 2018). Measurements performed at various *z*-positions of the objective do not evidence significant differences of the PL decay curves (the estimated lifetime being around 2 ns for all *z* values), ruling out potential effects related to self-absorption and re-emission (Figure S7 of the Supporting Information). A similar behaviour of the lifetime was reported also for other acenes, such as anthracene and tetracene (Ahn *et al.* 2008 and Camposeo *et al.* 2010), and was attributed to excitonic and defect or trapped states. Overall, the results of the microscopic PL measurements are consistent with the crystallographic analysis and support the different molecular packing of the investigated compounds in crystalline samples. In fact, F4 TMS ANT crystallizes as elongated needle-like samples, with parallel offset $\pi$-$\pi$ interaction and columnar stacking of chromophores. These configurations might lead to excimers (Liu *et al.* 2016) with resulting red-shifted, broad, long-lasting emission. On the contrary, the increased separation of the molecules in crystals of TIPS ANT*p* and the weaker molecular interactions, allow the electronic properties of the individual molecules to be partially preserved in the crystalline samples, which show structured PL spectra ascribable to different emitting species.

Crystallographic data of the two crystal structures have been deposited at the Cambridge Crystallographic Data Centre (CCDC) with deposit number CCDC1962253 for F4 TMS ANT and CCDC1962254 for TIPS ANT*p* and can be obtained free of charge *via* https://www.ccdc.cam.ac.uk/structures/.

*Acknowledgements*. A. C. and D. P. acknowledge funding from the European Research Council under the European Union's Horizon 2020 Research and Innovation Programme (Grant Agreement n. 682157, "*x*PRINT") and from the Italian Minister of University and Research under the PRIN program (grant no. 201795SBA3 "HARVEST"). J.A. thanks the U.S. National Science Foundation under cooperative agreement No. 1849213.

**4. Conclusions**

The crystal structure of two silylethyne-substituted anthracene compounds [*i.e*., 1,2,3,4-Tetrafluoro-5,8-bis(trimethylsilylethynyl)anthracene and a new polymorph of 9,10-bis(triisopropylsilylethynyl)anthracene] was determined by single-crystal synchrotron diffraction to identify main factors influencing the optical properties of these organic semiconductors. The crystallographic study revealed that the two compounds were characterized by different intermolecular interactions, responsible for dissimilar luminescence effects. The crystal morphology also affects the optical properties of the two compounds: needle-shape crystals are characterized by $\pi-\pi$ interactions and show broad and long-lasting photoluminescence. This was not found in case of the second crystal that was platelet shaped, for which a brighter PL with shorter lifetime was also observed, that could be exploited for the development of efficient light-emitting components and optical sensors.

**Table 1** Experimental and refinement details for F4 TMS ANT [present work and published results by Camposeo et al. (2019)] and TIPS ANTp.

|  | F4 TMS ANT present work | F4 TMS ANT (Camposeo et al., 2019) | TIPS ANTp |
|---|---|---|---|
| *Crystal data* | | | |
| Chemical formula | $C_{24}H_{22}F_4Si_2$ | $C_{24}H_{22}F_4Si_2$ | $C_{36}H_{50}Si_2$ |
| $M_r$ | 442.59 | 442.59 | 538.94 |
| Crystal system, space group | Monoclinic, $P2_1/c$ | Monoclinic, $P2_1/c$ | Triclinic, $P\text{-}1$ |
| Temperature (K) | 296 | 90 | 296 |
| $a, b, c$ (Å) | 6.9050 (14), 14.948 (3), 23.660 (5) | 6.7352 (4), 14.8514 (8), 23.3648 (13) | 8.6020 (17), 10.085 (2), 11.209 (2) |
| $\alpha, \beta, \gamma$ (°) | 90, 94.82 (3), 90 | 90, 94.735 (4), 90 | 115.07 (3), 102.61 (3), 98.82 (3) |
| $V$ (Å$^3$) | 2433.5 (8) | 2329.1 (2) | 825.6 (4) |
| $Z$ | 4 | 4 | 1 |
| Radiation type | Synchrotron, $\lambda$=0.72932 Å | Cu $K_\alpha$ | Synchrotron, $\lambda$=0.72932 Å |
| Crystal size (mm) | 0.09 × 0.01 × 0.01 | 0.23 × 0.05 × 0.05 | 0.09 × 0.077 × 0.077 |
| $\mu$ (mm$^{-1}$) | 0.19 | 1.74 | 0.14 |
| *Data collection* | | | |
| Diffractometer | Multi-axis PRIGo goniometer | Bruker X8 Proteum diffractometer | Multi-axis PRIGo goniometer |
| No. of measured, independent and observed [$I > 2\sigma(I)$] reflections | 62204, 4807, 3237 | 4199, 4199, 4036 | 27676, 4620, 4447 |
| $R_{int}$ | 0.059 | ------ | 0.026 |
| $(\sin\theta/\lambda)_{max}$ (Å$^{-1}$) | 0.617 | 0.603 | 0.716 |
| *Refinement* | | | |
| Bond precision (C−C) (Å) | 0.0032 | 0.0061 | 0.0018 |
| $R[F^2 > 2\sigma(F^2)]$, $wR(F^2)$, $S$ | 0.048, 0.148, 1.02 | 0.062, 0.175, 1.05 | 0.039, 0.116, 1.06 |
| No. of reflections | 4807 | 4303 | 4620 |
| No. of parameters | 277 | 279 | 178 |
| H-atom treatment | H-atom parameters constrained | H-atom parameters constrained | H-atom parameters constrained |
| $\Delta\rho_{max}, \Delta\rho_{min}$ (e Å$^{-3}$) | 0.22, −0.32 | 0.48, −0.45 | 0.32, −0.32 |

Computer programs: *XDS* (Kabsch, 2010), *SIR2019* (Burla *et al*., 2015), *SHELXL2014*/7 (Sheldrick, 2015), *WinGX* (Farrugia, 2012), *publCIF* (Westrip, 2010), *Mercury* (Macrae *et al*., 2008) and *CrystalExplorer17* (Turner *et al*., 2017).

**References**

Ahn, T.-S., Müller, A. M., Al-Kaysi, R.O., Spano, F.C., Norton, J.E., Beljonne, D., Bredas, J.-L., Bardeen, C.J. (2008). *J. Chem. Phys*. **128**, 054505.

Allard, S., Forster, M., Souharce, B., Thiem, H., Scherf, U. (2008). *Angew. Chem. Int. Ed.* **25**, 4070–4098.

Anthony, J.E. (2006). *Chem. Rev.* **106**, 5028–5048.

Anthony, J.E. (2008). *Angew. Chem. Int. Ed.* **47**, 452–483.

Anthony, J.E., Eaton, D.L., Parkin, S.R. (2002). *Org. Lett*. **4**, 15–18.

Anthony, J.E., Parkin, S.R. (2016). *CSD Communication.*

Bhattacharryya, K., Datta, A. (2017). *J. Phys. Chem. C* **121**, 1412-1420.

Burla, M.C., Caliandro, R., Carrozzini, B., Cascarano, G.L., Cuocci, C., Giacovazzo, C., Mallamo, M. Mazzone, A., Polidori, G. (2015). *J. Appl. Cryst*. **48**, 306-309.

Camposeo, A., Granger, D.B., Parkin, S.R., Altamura, D., Giannini, C., Anthony, J.E., Pisignano, D. (2019). *Chem. Mater*. **31**, 1775-1783.

Camposeo, A. Polo, M., Tavazzi, S., Silvestri, L., Spearman, P., Cingolani, R., Pisignano, D. (2010). *Phys. Rev. B*, **81**, 033306.

Chen, Z., Muller, P., Swager, T.M. (2006). *Org. Lett*. **8**, 273–276.

Da Silva Filho, D. A., Kim, E.-G., Brédas J.-L. (2005). *Adv. Mater.* **17**, 1072–1076.

del Valle, J.C., Turek, A.M., Tarkalanov, N.D., Saltiel, J. (2002). *J. Phys. Chem.* A**106**, 5101–5104.

Duo, L., You, J., Hong, Z., Xu, Z., Li, G., Street, R.A., Yang, Y. (2013). *Adv. Mater.* **25**, 6642–6671.

Farrugia, L.J. (2012). *J. Appl. Cryst*. **45**, 849–854.

Giacovazzo, C. (2014). *Phasing in Crystallography: A Modern Perspective*. Oxford University Press.

Griffith, O.L., Jones, A.G., Anthony, J.E., Lichtenberger, D.L. (2010). *J. Phys. Chem.* C **114**, 13838–13845.

Gu, X , Yao, J., Zhang, G., Yan, Y., Zhang, C., Peng, Q., Liao, Q., Wu, Y., Xu, Z., Zhao, Y., Fu, H., Zhang, D. (2012). *Adv. Funct. Mater*. **22**, 4862–4872.

Hattab, Z., Barbey, C., Monteil, M., Retailleau, P., Aouf, N.-E., Lecouvey, M., Dupont N. (2010). *J. Mol. Struct*. **973**, 144-151.

Ito, K., Suzuki, T., Sakamoto, Y., Kubota, D., Inoue, Y., Sako, F., Tokito, S. (2003). *Angew. Chem. Int. Ed*. **42**, 1159–1162.

Kabsch, W. (2010). *Acta Cryst*. D**66**, 125–132.

Kena-Cohen, S., Forrest, S. R. (2010). *Nat. Photonics* **4**, 371-375.

Kovalski, E., Korb, M., Hildebrandt, A. (2008). *Eur. J. Inorg. Chem.* **41**, 617-675.

Janiak, C. (2000). *J. Chem. Soc., Dalton Trans.* 3885–3896.

Landis, C.A., Parkin S.R., Anthony J.E. (2005). *Jpn. J. Appl. Phys.* **44**, 3921–3922.

Liu, Z., Xiao, J., Fu, Q., Feng, H., Zhang, X., Ren, T., Wang, S., Ma, D., Wang, X., Chen, H. (2013). *ACS Appl. Mater. Interfaces* **5**, 11136–11141.

Liu, H., Yao, L., Li, B., Chen, X., Gao, Y., Zhang, S., Li., W., Lu, P., Yang, B., Ma, Y. (2016). *Chem. Commun.* **52**, 7356-7359.

Macrae, C.F., Bruno, I.J., Chisholm, J.A., Edgington, P.R., McCabe, P., Pidcock, E., Rodriguez-Monge, L., Taylor, R. , van de Streek, J., Wood, P. A. (2008). *J. Appl. Cryst.* **41**, 466-470.

Martinez, C.R., Iverson, B.L. (2012). *Chem. Sci*. **3**, 2191–2201.

Mei, J., Diao, Y., Appleton, A.L., Fang, L., Bao, Z. (2013). *J. Am. Chem. Soc*. **135**, 6724–6746.

Meyer, E.A., Castellano, R.K., Diederich, F. (2003). *Angew. Chem. Int. Ed*. **42**, 1210-1250.

Nishio, M. (2004). *CrystEngComm* **6**, 130–158.

Nishio, M., Umezawa, Y., Suezawa, H., Tsuboyama, S. (2012). *The CH/π Hydrogen Bond: Implication in Crystal Engineering* In *The Importance of Pi-Interactions in Crystal Engineering: Frontiers in Crystal Engineering*, First Edition. Ed. By E.R.T. Tiekink and J. Zukermam-Schpector. John Wiley & Sons.

Ostroverkhova, O. (2016). *Chem. Rev*. **116**, 13279–13412.

Palilis, L.C., Lane, P.A., Kushto, G.P., Purushothaman, B., Anthony, J.E., Hafafi, Z.H. (2008). *Org. Electron*. **9**, 747–752.

Pun, J.K.H., Gallaher, J.K., Frazer, L., Prasad, S.K.K., Dover, C. B., MacQueen, R. W., Timothy W. Schmidt, T. W. (2018) *J. Photon. Energy* **8**, 022006.

Sheldrick, G.M. (2015). *Acta Cryst*. A**71**, 3–8.

Swenberg, C.E., Pope, M. (1999). *Electronic Processes of Organic Crystals*. G.M. (2015).Oxford, University Press, Oxford, NY.

Teka, S., Hriz, K., Jaballah, N. Kreher, D., Mathevet, F., Jarroux, N., Majdoub, M. (2015). *Materials Science in Semiconductor Processing* **34**, 189–197.

Turner, M.J., McKinnon, J.J., Wolff, S.K., Grimwood, D.J., Spackman, P.R., Jayatilaka, D., Spackman, M.A. *CrystalExplorer17* (2017). University of Western Australia. http://hirshfeldsurface.net.

Waltersperger, S., Olivier, V., Pradervand, C., Glettig, W, Salathe, M., Fuchs, M.R., Curtin, A., Wang, X., Ebner, S., Panepucci, E., Weinert, T., Schulze-Briese, C., Wang, M. (2015). *J. Synchrotron Rad*. **22**, 895-900.

Wang, S., Kappl, M., Liebewirth, I., Müller, M., Kirchoff, K., Pisula, W., Müllen K. (2012). *Adv. Mater*. **24**, 417–420.

Westrip, S.P. (2010). *J. Appl. Cryst*. **43**, 920–925.

Yao, Z.-F., Wang, J.-Y., Pei, J. (2018). *Cryst. Growth Des*. **18**, 7–15.

*Highly Efficient OLEDs with Phosphorescent Materials.* (2008). Ed. by H. Yersin, WILEY-VCH Verlag GmbH & Co. KGaA, Weinheim.

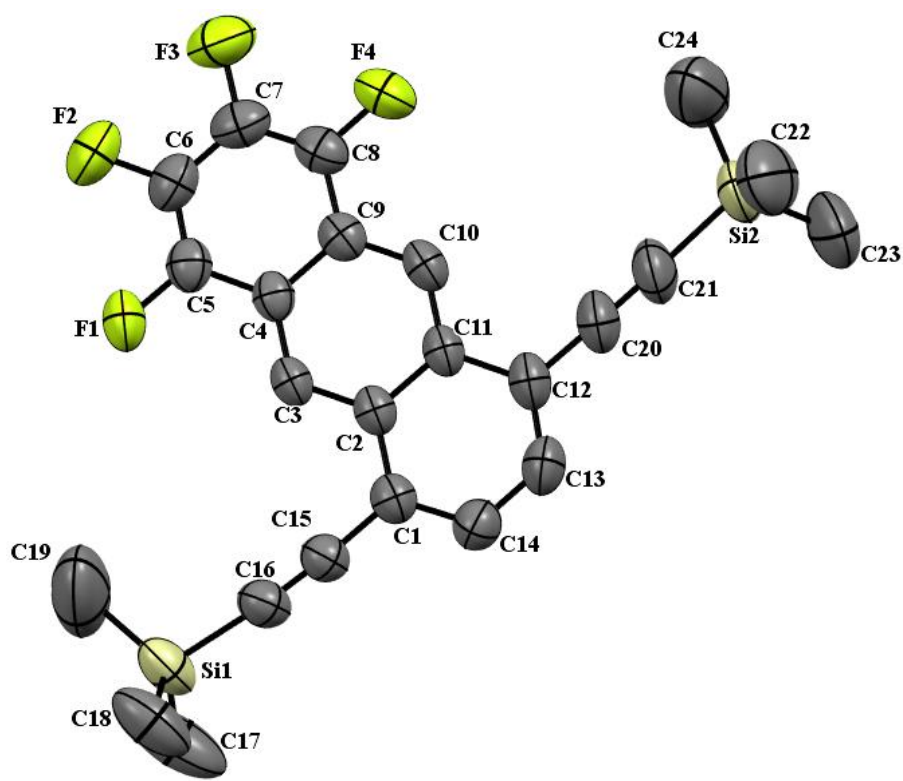

**Figure 1** F4 TMS ANT: A view along *a* of the asymmetric unit with the atomic labelling scheme. H atoms are omitted for clarity. Ellipsoids are drawn at 50% of probability level.

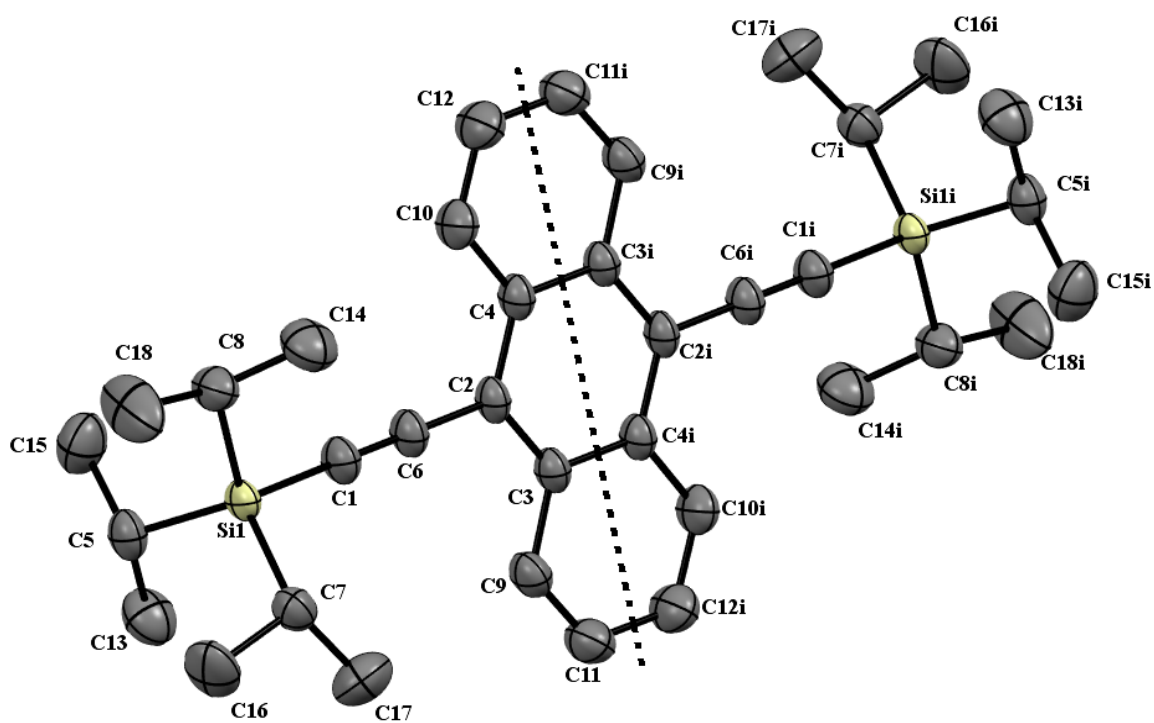

**Figure 2** TIPS ANT*p*: a view of the asymmetric unit (half of molecule, drawn at the left side of the broken line), showing the atom labelling scheme. H atoms are omitted for clarity. Ellipsoids are drawn at 50% of probability level [Symmetry code: (i) -*x*, 1-*y*, -*z*].

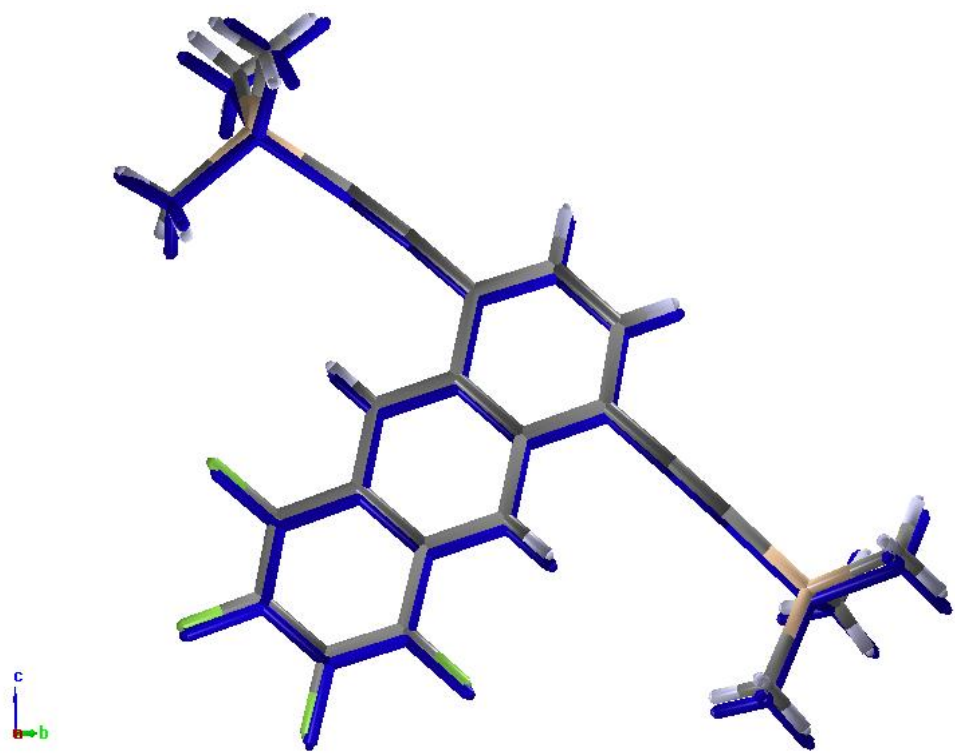

**Figure 3** F4 TMS ANT: A view of the structural overlay of the refined crystal structures, in case of the present work and of the published structure by Camposeo *et al.* (this last one is shown in blue colour) having an r.m.s. deviation of 0.04 Å (*SIR2019*; Burla *et al.*, 2015).

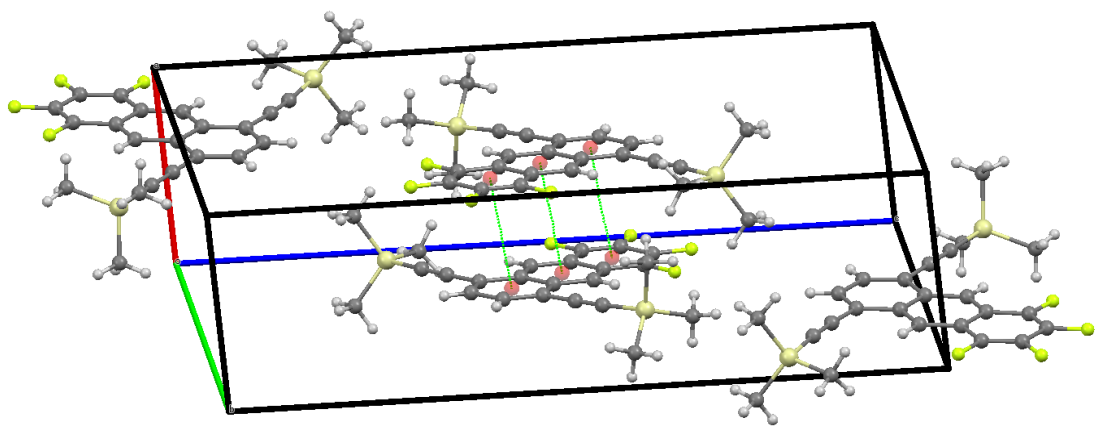

**Figure 4** F4 TMS ANT: a view of crystal packing (*a* axis, in red) showing the parallel-offset π-π interactions that are indicated by green broken lines between centroids of the aromatic rings (represented with red spheres).

**Figure 5** F4 TMS ANT: a view of the local environment of the molecule represented with sticks and its neighbouring entities (represented with wireframes); centroids are represented with red spheres. The $\pi$-$\pi$ interactions are indicated by broken red lines while $C_{aryl}-F \cdots H-C$ interactions by broken blue lines.

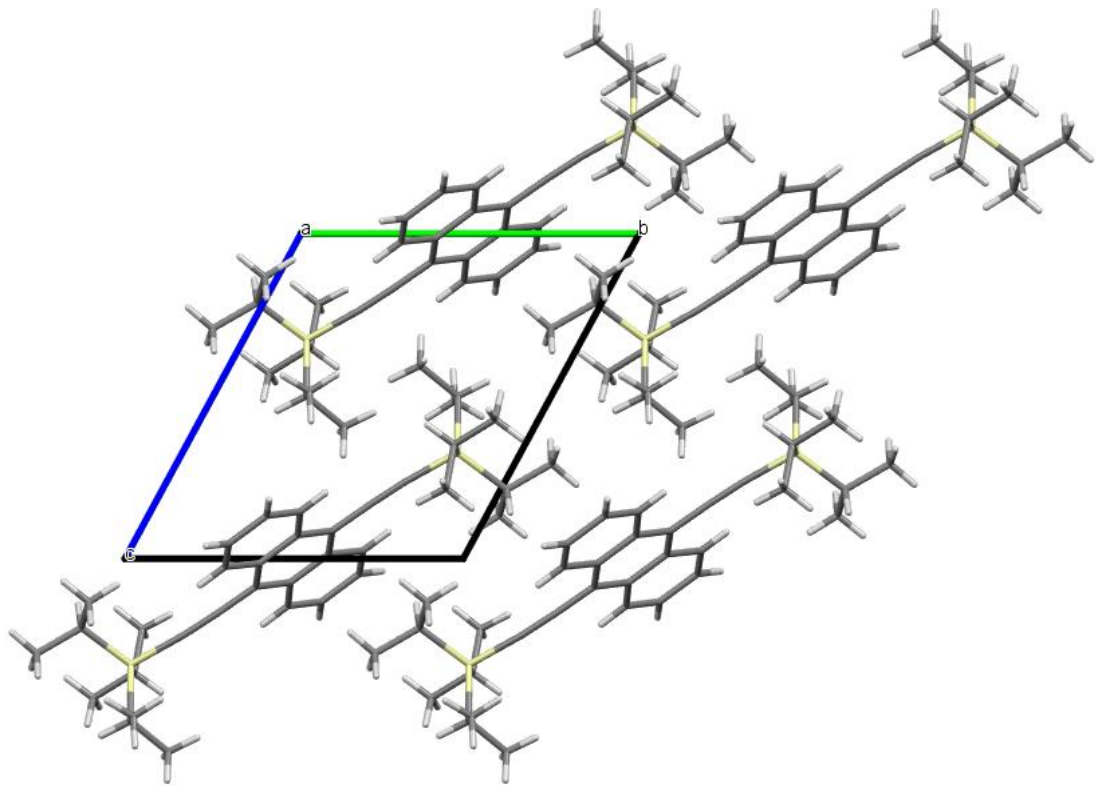

**Figure 6** TIPS ANT*p*: a view along *a* of the crystal packing.

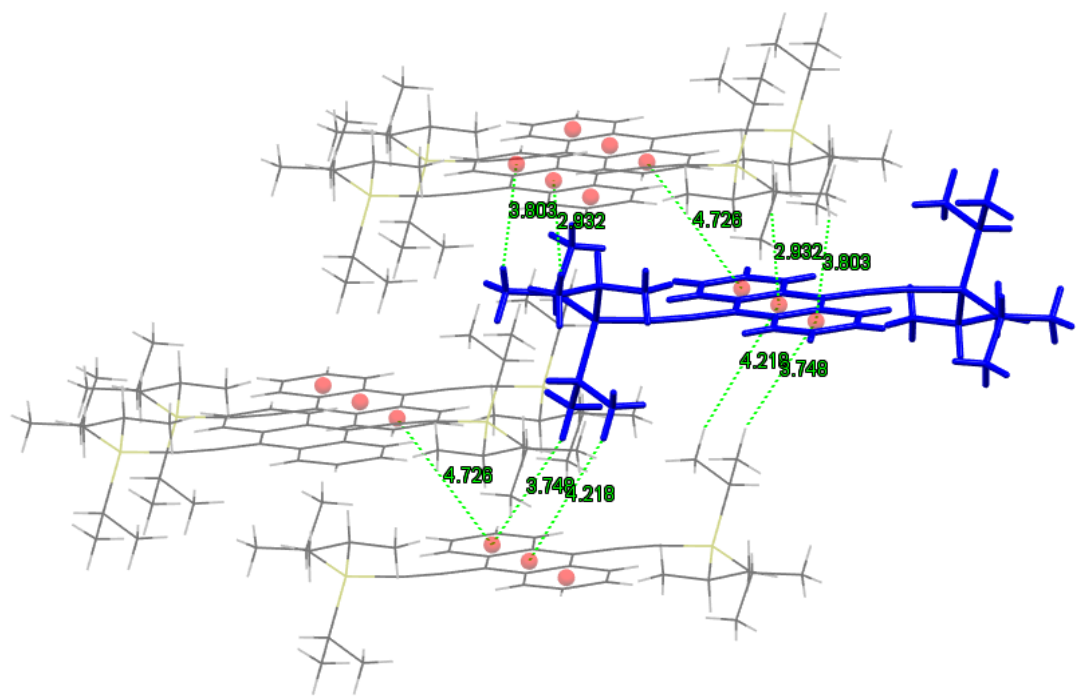

**Figure 7** TIPS ANT*p*: a view of the local environment of the molecule represented with sticks and its neighbouring entities (represented with wireframes); centroids are represented with red spheres. The weak π-π interactions (with a distance between centroids of 4.726 Å) are indicated by broken green lines as well as the C-H···π interactions.

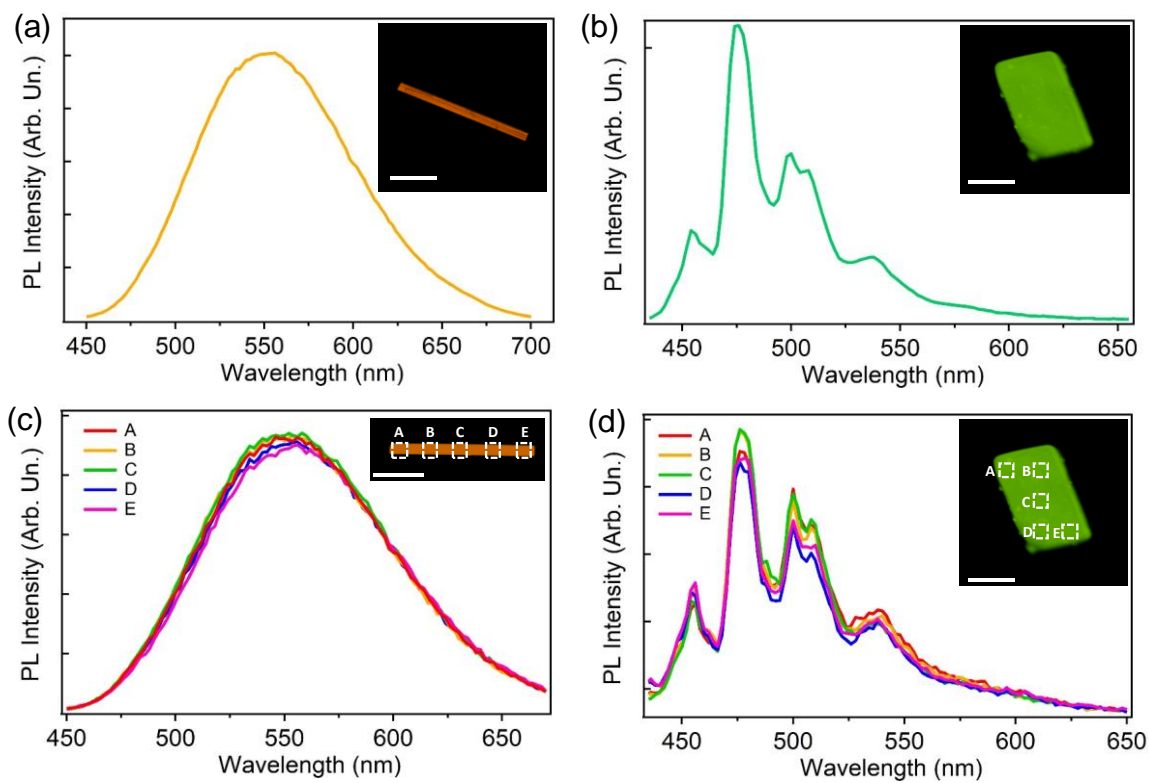

**Figure 8** (a)-(b) PL spectra of F4 TMS ANT (a) and TIPS ANT*p* (b), measured by confocal microscopy. The corresponding micrographs of the PL intensity of the investigated samples are shown in the insets. Inset scale bars: 50 μm. (c)-(d) Comparison of PL spectra of F4 TMS ANT (c) and TIPS ANT*p* (d), measured by confocal microscopy in various regions of the crystalline samples. The corresponding insets show confocal PL micrographs of the investigated samples. Scale bars: 50 μm. The spectra (A-E) are measured in the areas marked with the dashed squares in the insets. The spectra shown in (c) and (d) and the PL micrographs were acquired by using the objective with NA=0.4.

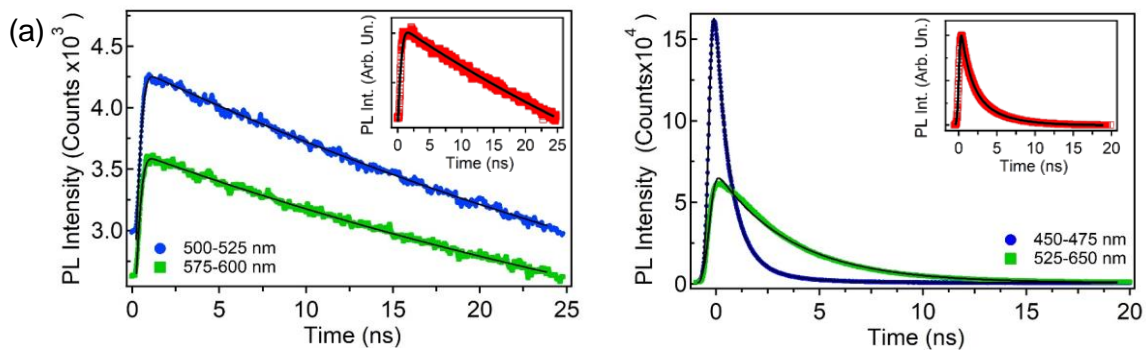

**Figure 9**. Decay of the PL intensity of F4 TMS ANT (a) and TIPS ANT*p* (b), in different spectral ranges. The insets of (a) and (b) display the decay of the PL intensity integrated in the whole emission range. The continuous black lines are fits to data by exponential functions convoluted with the instrumental response function.

**Supporting Information**

# Synthesis, crystal structure, polymorphism and microscopic luminescence properties of anthracene derivative compounds

**Anna Moliterni[a]\*, Davide Altamura[a], Rocco Lassandro[a], Vincent Olieric[b], Gianmarco Ferri[c], Francesco Cardarelli[c], Andrea Camposeo[d], Dario Pisignano[de], John E. Anthony[f] and Cinzia Giannini[a]\***

[a]Istituto di Cristallografia, CNR, Via Amendola, 122/O, Bari, 70126, Italy

[b] Paul Scherrer Institute, Forschungstrasse 111, Villigen-PSI, 5232, Switzerland

[c]NEST, Scuola Normale Superiore, Piazza San Silvestro 12, Pisa, I-56127, Italy

[d]NEST, Istituto Nanoscienze, CNR, Piazza San Silvestro 12, Pisa, I-56127, Italy

[e]Dipartimento di Fisica 'E. Fermi', University of Pisa, Pisa, I-56127, Italy

[f]Center for Applied Energy Research, University of Kentucky, Research Park Drive, Lexington, KY, 2582, USA

Correspondence email: annagrazia.moliterni@ic.cnr.it; cinzia.giannini@ic.cnr.it

Cell refinement: *XDS* (Kabsch, 2010); data reduction: *XDS* (Kabsch, 2010); program used to solve structure: *SIR2019* (Burla *et al.*, 2015); program used to refine structure: *SHELXL2014*/7 (Sheldrick, 2014); molecular graphics: *SIR2019* (Burla *et al.*, 2015), Mercury (Macrae *et al.*, 2008); software used to prepare material for publication: *WinGX* (Farrugia, 2012) and *publCIF* (Westrip, 2010).

**F4 TMS ANT**

*Crystal data*

| | |
|---|---|
| $C_{24}H_{22}F_4Si_2$ | $F(000) = 920$ |
| $M_r = 442.59$ | $D_x = 1.208$ Mg m$^{-3}$ |
| Monoclinic, $P2_1/c$ | Synchrotron radiation, $\lambda = 0.72932$ Å |
| $a = 6.9050\,(14)$ Å | Cell parameters from 421 reflections |
| $b = 14.948\,(3)$ Å | $\theta = 1.7–26.8°$ |
| $c = 23.660\,(5)$ Å | $\mu = 0.194$ mm$^{-1}$ |
| $\beta = 94.82\,(3)°$ | $T = 296$ K |
| $V = 2433.5\,(8)$ Å$^3$ | Needle, yellow |
| $Z = 4$ | $0.09 \times 0.01 \times 0.01$ mm |

*Data collection*

| | |
|---|---|
| Multi-axis PRIGo goniometer diffractometer | $R_{int} = 0.0592$ |
| Radiation source: Synchrotron, beamline PXIII, Swiss Light Source (SLS), Villigen, Switzerland | $\theta_{max} = 26.8°$, $\theta_{min} = 1.7°$ |
| $\omega$ scans, shutterless continuous rotation method | $h = -8 \;\; 8$ |
| 62204 measured reflections | $k = -18 \;\; 18$ |
| 4807 independent reflections | $l = -29 \;\; 29$ |
| 3237 reflections with $I > 2\sigma(I)$ | |

*Refinement*

| | |
|---|---|
| Refinement on $F^2$ | Primary atom site location: structure-invariant direct methods |
| Least-squares matrix: full | Hydrogen site location: inferred from neighbouring sites |
| $R[F^2 > 2\sigma(F^2)] = 0.048$ | H-atom parameters constrained |
| $wR(F^2) = 0.148$ | $w = 1/[\sigma^2(F_o^2) + (0.0702P)^2 + 0.5392P]$ where $P = (F_o^2 + 2F_c^2)/3$ |
| $S = 1.02$ | $(\Delta/\sigma)_{max} = 0.001$ |
| 4807 reflections | $\Delta\rho_{max} = 0.22$ e Å$^{-3}$ |
| 277 parameters | $\Delta\rho_{min} = -0.32$ e Å$^{-3}$ |
| 0 restraints | |

*Special details*

*Geometry*. All esds (except the esd in the dihedral angle between two l.s. planes) are estimated using the full covariance matrix. The cell esds are taken into account individually in the estimation of esds in distances, angles and torsion angles; correlations between esds in cell parameters are only used when they are defined by crystal symmetry. An approximate (isotropic) treatment of cell esds is used for estimating esds involving l.s. planes.

*Refinement*. Refinement of $F^2$ against ALL reflections. The weighted *R*-factor *wR* and goodness of fit *S* are based on $F^2$, conventional *R*-factors *R* are based on *F*, with *F* set to zero for negative $F^2$. The threshold expression of $F^2 > 2\sigma(F^2)$ is used only for calculating *R*-factors(*gt*) etc. and is not relevant to the choice of reflections for refinement. *R*-factors based on $F^2$ are statistically about twice as large as those based on *F*, and *R*-factors based on ALL data will be even larger.

*Fractional atomic coordinates and isotropic or equivalent isotropic displacement parameters (Å$^2$)*

|      | x              | y             | z              | $U_{iso}$*/$U_{eq}$ |
|------|----------------|---------------|----------------|---------------------|
| Si1  | 0.77925 (12)   | 0.84835 (5)   | 0.22633 (4)    | 0.0897 (3)          |
| Si2  | 0.64819 (16)   | 1.45054 (5)   | −0.09270 (4)   | 0.1057 (3)          |
| F1   | 0.7797 (2)     | 0.80384 (8)   | −0.00591 (6)   | 0.0807 (4)          |
| F2   | 0.7872 (2)     | 0.77163 (10)  | −0.11865 (7)   | 0.1013 (5)          |
| F3   | 0.7719 (3)     | 0.90604 (13)  | −0.19354 (6)   | 0.1106 (6)          |
| F4   | 0.7492 (2)     | 1.07801 (10)  | −0.15848 (6)   | 0.0898 (4)          |
| C1   | 0.7337 (3)     | 1.08694 (14)  | 0.11365 (9)    | 0.0595 (5)          |
| C2   | 0.7382 (3)     | 1.06734 (12)  | 0.05430 (8)    | 0.0527 (4)          |
| C3   | 0.7522 (3)     | 0.97987 (12)  | 0.03488 (8)    | 0.0536 (5)          |
| H3   | 0.7581         | 0.9331        | 0.0609         | 0.064*              |
| C4   | 0.7576 (3)     | 0.96084 (12)  | −0.02229 (9)   | 0.0539 (5)          |
| C5   | 0.7715 (3)     | 0.87214 (14)  | −0.04318 (10)  | 0.0636 (5)          |
| C6   | 0.7754 (3)     | 0.85560 (16)  | −0.09880 (12)  | 0.0739 (6)          |
| C7   | 0.7679 (4)     | 0.92591 (19)  | −0.13812 (10)  | 0.0771 (6)          |
| C8   | 0.7559 (3)     | 1.01123 (17)  | −0.12058 (9)   | 0.0685 (6)          |
| C9   | 0.7493 (3)     | 1.03292 (14)  | −0.06245 (9)   | 0.0569 (5)          |
| C10  | 0.7343 (3)     | 1.12031 (13)  | −0.04307 (9)   | 0.0595 (5)          |
| H10  | 0.7281         | 1.1671        | −0.0691        | 0.071*              |
| C11  | 0.7283 (3)     | 1.13933 (12)  | 0.01422 (9)    | 0.0558 (5)          |
| C12  | 0.7119 (3)     | 1.22989 (13)  | 0.03469 (10)   | 0.0669 (6)          |
| C13  | 0.7098 (4)     | 1.24464 (15)  | 0.09152 (12)   | 0.0770 (6)          |
| H13  | 0.7008         | 1.3031        | 0.1045         | 0.092*              |
| C14  | 0.7207 (3)     | 1.17456 (15)  | 0.13110 (11)   | 0.0717 (6)          |
| H14  | 0.7193         | 1.1874        | 0.1695         | 0.086*              |
| C15  | 0.7448 (3)     | 1.01472 (15)  | 0.15357 (9)    | 0.0633 (5)          |
| C16  | 0.7560 (3)     | 0.95084 (16)  | 0.18430 (10)   | 0.0714 (6)          |
| C17  | 0.9936 (5)     | 0.8601 (3)    | 0.27760 (17)   | 0.1577 (19)         |
| H17A | 1.0110         | 0.8063        | 0.2996         | 0.237*              |
| H17B | 0.9748         | 0.9097        | 0.3024         | 0.237*              |
| H17C | 1.1068         | 0.8705        | 0.2576         | 0.237*              |
| C18  | 0.5596 (4)     | 0.8325 (3)    | 0.26395 (16)   | 0.1257 (13)         |
| H18A | 0.5786         | 0.7832        | 0.2898         | 0.189*              |
| H18B | 0.4511         | 0.8203        | 0.2370         | 0.189*              |
| H18C | 0.5345         | 0.8859        | 0.2847         | 0.189*              |
| C19  | 0.8126 (9) | 0.7584 (3)   | 0.1759 (2)    | 0.200 (3)   |
| H19A | 0.8515     | 0.7049       | 0.1962        | 0.300*      |
| H19B | 0.9112     | 0.7752       | 0.1517        | 0.300*      |
| H19C | 0.6926     | 0.7479       | 0.1534        | 0.300*      |
| C20  | 0.6958 (4) | 1.30201 (14) | −0.00519 (12) | 0.0808 (7)  |
| C21  | 0.6799 (5) | 1.36126 (16) | −0.03916 (14) | 0.0945 (8)  |
| C22  | 0.3877 (6) | 1.4793 (3)   | −0.10196 (18) | 0.1435 (15) |
| H22A | 0.3630     | 1.5156       | −0.1353       | 0.215*      |
| H22B | 0.3524     | 1.5118       | −0.0694       | 0.215*      |
| H22C | 0.3121     | 1.4254       | −0.1061       | 0.215*      |
| C23  | 0.7880 (8) | 1.5485 (2)   | −0.06689 (19) | 0.1564 (17) |
| H23A | 0.7752     | 1.5950       | −0.0950       | 0.235*      |
| H23B | 0.9224     | 1.5325       | −0.0598       | 0.235*      |
| H23C | 0.7396     | 1.5695       | −0.0324       | 0.235*      |
| C24  | 0.7293 (8) | 1.4056 (3)   | −0.15961 (18) | 0.1625 (18) |
| H24A | 0.6947     | 1.4468       | −0.1900       | 0.244*      |
| H24B | 0.6676     | 1.3490       | −0.1679       | 0.244*      |
| H24C | 0.8678     | 1.3978       | −0.1557       | 0.244*      |

*Atomic displacement parameters (Å²)*

|     | $U^{11}$    | $U^{22}$    | $U^{33}$    | $U^{12}$     | $U^{13}$    | $U^{23}$     |
|-----|-------------|-------------|-------------|--------------|-------------|--------------|
| Si1 | 0.0933 (5)  | 0.0888 (5)  | 0.0896 (5)  | 0.0186 (4)   | 0.0234 (4)  | 0.0364 (4)   |
| Si2 | 0.1574 (9)  | 0.0561 (4)  | 0.1054 (6)  | 0.0177 (4)   | 0.0213 (6)  | 0.0216 (4)   |
| F1  | 0.0933 (9)  | 0.0467 (7)  | 0.1017 (10) | 0.0006 (6)   | 0.0063 (8)  | 0.0056 (6)   |
| F2  | 0.1177 (12) | 0.0757 (9)  | 0.1104 (12) | −0.0027 (8)  | 0.0082 (9)  | −0.0338 (8)  |
| F3  | 0.1314 (13) | 0.1274 (14) | 0.0730 (10) | −0.0060 (11) | 0.0087 (9)  | −0.0245 (10) |
| F4  | 0.1034 (10) | 0.1000 (10) | 0.0660 (8)  | 0.0003 (8)   | 0.0061 (7)  | 0.0221 (7)   |
| C1  | 0.0573 (11) | 0.0557 (11) | 0.0655 (12) | −0.0003 (9)  | 0.0051 (9)  | 0.0047 (9)   |
| C2  | 0.0470 (10) | 0.0470 (10) | 0.0638 (12) | 0.0002 (8)   | 0.0041 (8)  | 0.0076 (8)   |
| C3  | 0.0504 (10) | 0.0461 (10) | 0.0640 (12) | 0.0001 (8)   | 0.0034 (8)  | 0.0112 (9)   |
| C4  | 0.0465 (10) | 0.0480 (10) | 0.0666 (12) | −0.0018 (8)  | 0.0018 (8)  | 0.0044 (9)   |
| C5  | 0.0575 (12) | 0.0554 (11) | 0.0773 (14) | −0.0030 (9)  | 0.0017 (10) | 0.0001 (10)  |
| C6  | 0.0698 (14) | 0.0645 (13) | 0.0872 (17) | −0.0032 (11) | 0.0050 (12) | −0.0161 (12) |
| C7  | 0.0744 (15) | 0.0929 (18) | 0.0638 (14) | −0.0051 (13) | 0.0037 (11) | −0.0137 (13) |
| C8  | 0.0638 (13) | 0.0790 (15) | 0.0619 (13) | −0.0023 (11) | 0.0013 (10) | 0.0101 (11)  |
| C9  | 0.0483 (10) | 0.0593 (11) | 0.0626 (12) | −0.0019 (8)  | 0.0016 (8)  | 0.0070 (9)   |
| C10 | 0.0565 (11) | 0.0529 (11) | 0.0689 (13) | −0.0007 (9)  | 0.0031 (9)  | 0.0169 (10)  |
| C11 | 0.0525 (10) | 0.0447 (10) | 0.0698 (13) | −0.0010 (8)  | 0.0027 (9)  | 0.0084 (9)   |
| C12 | 0.0684 (13) | 0.0462 (11) | 0.0861 (16) | 0.0013 (9)   | 0.0059 (11) | 0.0076 (10)  |
| C13 | 0.0890 (16) | 0.0477 (11) | 0.0941 (18) | 0.0026 (11) | 0.0069 (13) | −0.0059 (11) |
| C14 | 0.0800 (15) | 0.0632 (13) | 0.0717 (14) | 0.0022 (11) | 0.0047 (11) | −0.0060 (11) |
| C15 | 0.0642 (12) | 0.0654 (12) | 0.0605 (12) | 0.0023 (10) | 0.0065 (9) | 0.0042 (10) |
| C16 | 0.0763 (15) | 0.0776 (15) | 0.0608 (13) | 0.0045 (11) | 0.0087 (11) | 0.0089 (11) |
| C17 | 0.089 (2) | 0.246 (5) | 0.138 (3) | 0.024 (3) | 0.009 (2) | 0.109 (3) |
| C18 | 0.089 (2) | 0.162 (3) | 0.129 (3) | 0.006 (2) | 0.0206 (18) | 0.077 (2) |
| C19 | 0.297 (7) | 0.089 (3) | 0.227 (6) | 0.027 (3) | 0.096 (5) | 0.005 (3) |
| C20 | 0.0915 (17) | 0.0479 (12) | 0.1032 (19) | 0.0036 (11) | 0.0092 (14) | 0.0097 (12) |
| C21 | 0.117 (2) | 0.0529 (13) | 0.114 (2) | 0.0075 (13) | 0.0113 (17) | 0.0174 (14) |
| C22 | 0.175 (4) | 0.105 (3) | 0.150 (3) | 0.050 (3) | 0.003 (3) | 0.017 (2) |
| C23 | 0.236 (5) | 0.078 (2) | 0.155 (4) | −0.031 (3) | 0.013 (3) | 0.034 (2) |
| C24 | 0.246 (5) | 0.113 (3) | 0.136 (3) | 0.037 (3) | 0.063 (3) | 0.010 (3) |

*Geometric parameters (Å, °)*

| | | | |
|---|---|---|---|
| Si1—C19 | 1.825 (5) | C10—H10 | 0.9300 |
| Si1—C16 | 1.826 (2) | C11—C12 | 1.445 (3) |
| Si1—C18 | 1.837 (3) | C12—C13 | 1.364 (3) |
| Si1—C17 | 1.842 (4) | C12—C20 | 1.431 (3) |
| Si2—C23 | 1.831 (4) | C13—C14 | 1.403 (3) |
| Si2—C21 | 1.841 (3) | C13—H13 | 0.9300 |
| Si2—C22 | 1.844 (4) | C14—H14 | 0.9300 |
| Si2—C24 | 1.849 (4) | C15—C16 | 1.199 (3) |
| F1—C5 | 1.347 (2) | C17—H17A | 0.9600 |
| F2—C6 | 1.345 (3) | C17—H17B | 0.9600 |
| F3—C7 | 1.347 (3) | C17—H17C | 0.9600 |
| F4—C8 | 1.340 (3) | C18—H18A | 0.9600 |
| C1—C14 | 1.378 (3) | C18—H18B | 0.9600 |
| C1—C15 | 1.432 (3) | C18—H18C | 0.9600 |
| C1—C2 | 1.437 (3) | C19—H19A | 0.9600 |
| C2—C3 | 1.392 (3) | C19—H19B | 0.9600 |
| C2—C11 | 1.432 (3) | C19—H19C | 0.9600 |
| C3—C4 | 1.386 (3) | C20—C21 | 1.195 (3) |
| C3—H3 | 0.9300 | C22—H22A | 0.9600 |
| C4—C5 | 1.421 (3) | C22—H22B | 0.9600 |
| C4—C9 | 1.435 (3) | C22—H22C | 0.9600 |
| C5—C6 | 1.341 (3) | C23—H23A | 0.9600 |
| C6—C7 | 1.402 (4) | C23—H23B | 0.9600 |
| C7—C8 | 1.346 (3) | C23—H23C | 0.9600 |
| C8—C9 | 1.418 (3) | C24—H24A | 0.9600 |
| C9—C10 | 1.391 (3) | C24—H24B | 0.9600 |
| C10—C11 | 1.389 (3) | C24—H24C | 0.9600 |
| | | | |
| C19—Si1—C16 | 105.79 (18) | C20—C12—C11 | 119.3 (2) |
| C19—Si1—C18 | 112.1 (2) | C12—C13—C14 | 122.2 (2) |
| C16—Si1—C18 | 109.48 (13) | C12—C13—H13 | 118.9 |
| C19—Si1—C17 | 111.3 (3) | C14—C13—H13 | 118.9 |
| C16—Si1—C17 | 107.91 (16) | C1—C14—C13 | 120.7 (2) |
| C18—Si1—C17 | 110.11 (17) | C1—C14—H14 | 119.6 |
| C23—Si2—C21 | 108.76 (17) | C13—C14—H14 | 119.6 |
| C23—Si2—C22 | 109.6 (2) | C16—C15—C1 | 176.0 (2) |
| C21—Si2—C22 | 108.01 (17) | C15—C16—Si1 | 175.4 (2) |
| C23—Si2—C24 | 112.8 (2) | Si1—C17—H17A | 109.5 |
| C21—Si2—C24 | 107.39 (16) | Si1—C17—H17B | 109.5 |
| C22—Si2—C24 | 110.2 (2) | H17A—C17—H17B | 109.5 |
| C14—C1—C15 | 121.3 (2) | Si1—C17—H17C | 109.5 |
| C14—C1—C2 | 119.58 (19) | H17A—C17—H17C | 109.5 |
| C15—C1—C2 | 119.09 (18) | H17B—C17—H17C | 109.5 |
| C3—C2—C11 | 119.21 (18) | Si1—C18—H18A | 109.5 |
| C3—C2—C1 | 121.45 (17) | Si1—C18—H18B | 109.5 |
| C11—C2—C1 | 119.34 (17) | H18A—C18—H18B | 109.5 |
| C4—C3—C2 | 121.55 (17) | Si1—C18—H18C | 109.5 |
| C4—C3—H3 | 119.2 | H18A—C18—H18C | 109.5 |
| C2—C3—H3 | 119.2 | H18B—C18—H18C | 109.5 |
| C3—C4—C5 | 122.62 (18) | Si1—C19—H19A | 109.5 |
| C3—C4—C9 | 119.30 (18) | Si1—C19—H19B | 109.5 |
| C5—C4—C9 | 118.07 (19) | H19A—C19—H19B | 109.5 |
| C6—C5—F1 | 119.9 (2) | Si1—C19—H19C | 109.5 |
| C6—C5—C4 | 121.4 (2) | H19A—C19—H19C | 109.5 |
| F1—C5—C4 | 118.7 (2) | H19B—C19—H19C | 109.5 |
| C5—C6—F2 | 121.4 (2) | C21—C20—C12 | 178.7 (3) |
| C5—C6—C7 | 120.7 (2) | C20—C21—Si2 | 178.0 (3) |
| F2—C6—C7 | 117.9 (2) | Si2—C22—H22A | 109.5 |
| C8—C7—F3 | 121.1 (2) | Si2—C22—H22B | 109.5 |
| C8—C7—C6 | 120.4 (2) | H22A—C22—H22B | 109.5 |
| F3—C7—C6 | 118.6 (2) | Si2—C22—H22C | 109.5 |
| F4—C8—C7 | 120.0 (2) | H22A—C22—H22C | 109.5 |
| F4—C8—C9 | 118.5 (2) | H22B—C22—H22C | 109.5 |
| C7—C8—C9 | 121.6 (2) | Si2—C23—H23A | 109.5 |
| C10—C9—C8 | 122.95 (19) | Si2—C23—H23B | 109.5 |
| C10—C9—C4 | 119.14 (19) | H23A—C23—H23B | 109.5 |
| C8—C9—C4 | 117.91 (19) | Si2—C23—H23C | 109.5 |
| C11—C10—C9 | 121.53 (18) | H23A—C23—H23C | 109.5 |
| C11—C10—H10 | 119.2 | H23B—C23—H23C | 109.5 |
| C9—C10—H10 | 119.2 | Si2—C24—H24A | 109.5 |
| C10—C11—C2 | 119.26 (18) | Si2—C24—H24B | 109.5 |
| C10—C11—C12 | 121.89 (18) | H24A—C24—H24B | 109.5 |
| C2—C11—C12 | 118.86 (19) | Si2—C24—H24C | 109.5 |
| C13—C12—C20 | 121.4 (2) | H24A—C24—H24C | 109.5 |
| C13—C12—C11 | 119.3 (2) | H24B—C24—H24C | 109.5 |
| | | | |
| C14—C1—C2—C3 | 179.66 (19) | C7—C8—C9—C10 | −179.1 (2) |
| C15—C1—C2—C3 | 0.3 (3) | F4—C8—C9—C4 | −179.70 (17) |
| C14—C1—C2—C11 | −0.3 (3) | C7—C8—C9—C4 | 0.5 (3) |
| C15—C1—C2—C11 | −179.63 (18) | C3—C4—C9—C10 | −0.6 (3) |
| C11—C2—C3—C4 | 0.3 (3) | C5—C4—C9—C10 | 179.53 (18) |
| C1—C2—C3—C4 | −179.67 (17) | C3—C4—C9—C8 | 179.75 (18) |
| C2—C3—C4—C5 | −179.87 (18) | C5—C4—C9—C8 | −0.1 (3) |
| C2—C3—C4—C9 | 0.3 (3) | C8—C9—C10—C11 | −179.96 (19) |
| C3—C4—C5—C6 | 179.7 (2) | C4—C9—C10—C11 | 0.4 (3) |
| C9—C4—C5—C6 | −0.5 (3) | C9—C10—C11—C2 | 0.1 (3) |
| C3—C4—C5—F1 | 0.0 (3) | C9—C10—C11—C12 | −179.77 (19) |
| C9—C4—C5—F1 | 179.88 (17) | C3—C2—C11—C10 | −0.4 (3) |
| F1—C5—C6—F2 | 0.1 (3) | C1—C2—C11—C10 | 179.48 (17) |
| C4—C5—C6—F2 | −179.55 (19) | C3—C2—C11—C12 | 179.43 (18) |
| F1—C5—C6—C7 | −179.7 (2) | C1—C2—C11—C12 | −0.6 (3) |
| C4—C5—C6—C7 | 0.7 (4) | C10—C11—C12—C13 | −179.0 (2) |
| C5—C6—C7—C8 | −0.2 (4) | C2—C11—C12—C13 | 1.1 (3) |
| F2—C6—C7—C8 | 180.0 (2) | C10—C11—C12—C20 | 1.8 (3) |
| C5—C6—C7—F3 | −179.9 (2) | C2—C11—C12—C20 | −178.1 (2) |
| F2—C6—C7—F3 | 0.3 (3) | C20—C12—C13—C14 | 178.4 (2) |
| F3—C7—C8—F4 | −0.5 (3) | C11—C12—C13—C14 | −0.8 (4) |
| C6—C7—C8—F4 | 179.9 (2) | C15—C1—C14—C13 | −180.0 (2) |
| F3—C7—C8—C9 | 179.3 (2) | C2—C1—C14—C13 | 0.7 (3) |
| C6—C7—C8—C9 | −0.4 (4) | C12—C13—C14—C1 | −0.2 (4) |
| F4—C8—C9—C10 | 0.7 (3) | | |

**TIPS ANT$p$**

*Crystal data*

| | |
|---|---|
| $C_{36}H_{50}Si_2$ | $Z=1$ |
| $M_r = 538.94$ | $F(000) = 294$ |
| Triclinic, $P$-1 | $D_x = 1.084$ Mg m$^{-3}$ |
| $a = 8.6020$ (17) Å | Synchrotron radiation, $\lambda = 0.72932$ Å |
| $b = 10.085$ (2) Å | Cell parameters from 1335 reflections |
| $c = 11.209$ (2) Å | $\theta = 2.2$–$31.5°$ |
| $\alpha = 115.07$ (3)° | $\mu = 0.136$ mm$^{-1}$ |
| $\beta = 102.61$ (3)° | $T = 296$ K |
| $\gamma = 98.82$ (3)° | Block, yellow |
| $V = 825.6$ (4) Å$^3$ | $0.09 \times 0.08 \times 0.08$ mm |

*Data collection*

| | |
|---|---|
| Multi-axis PRIGo goniometer diffractometer | $R_{int} = 0.026$ |
| Radiation source: Synchrotron, beamline PXIII, Swiss Light Source (SLS), Villigen, Switzerland | $\theta_{max} = 31.5°$, $\theta_{min} = 2.2°$ |
| $\omega$ scans, shutterless continuous rotation method | $h = -12 \quad 12$ |
| 27676 measured reflections | $k = -14 \quad 14$ |
| 4620 independent reflections | $l = -15 \quad 16$ |
| 4447 reflections with $I > 2\sigma(I)$ | |

*Refinement*

| | |
|---|---|
| Refinement on $F^2$ | Primary atom site location: structure-invariant direct methods |
| Least-squares matrix: full | Hydrogen site location: inferred from neighbouring sites |
| $R[F^2 > 2\sigma(F^2)] = 0.039$ | H-atom parameters constrained |
| $wR(F^2) = 0.116$ | $w = 1/[\sigma^2(F_o^2) + (0.0706P)^2 + 0.1231P]$ where $P = (F_o^2 + 2F_c^2)/3$ |
| $S = 1.06$ | $(\Delta/\sigma)_{max} < 0.001$ |
| 4620 reflections | $\Delta\rho_{max} = 0.32$ e Å$^{-3}$ |
| 178 parameters | $\Delta\rho_{min} = -0.32$ e Å$^{-3}$ |
| 0 restraints | |

*Special details*

*Geometry*. All esds (except the esd in the dihedral angle between two l.s. planes) are estimated using the full covariance matrix. The cell esds are taken into account individually in the estimation of esds in distances, angles and torsion angles; correlations between esds in cell parameters are only used when they are defined by crystal symmetry. An approximate (isotropic) treatment of cell esds is used for estimating esds involving l.s. planes.

*Refinement*. Refinement of $F^2$ against ALL reflections. The weighted $R$-factor $wR$ and goodness of fit $S$ are based on $F^2$, conventional $R$-factors $R$ are based on $F$, with $F$ set to zero for negative $F^2$. The threshold expression of $F^2 > 2\sigma(F^2)$ is used only for calculating $R$-factors($gt$) etc. and is not relevant to the choice of reflections for refinement. *R*-factors based on $F^2$ are statistically about twice as large as those based on $F$, and *R*-factors based on ALL data will be even larger.

*Fractional atomic coordinates and isotropic or equivalent isotropic displacement parameters (Å$^2$)*

|      | *x*          | *y*          | *z*          | $U_{iso}$*/$U_{eq}$ |
|------|--------------|--------------|--------------|---------------------|
| Si1  | 0.18459 (3)  | 0.18680 (3)  | 0.32597 (2)  | 0.03045 (9)         |
| C1   | 0.12184 (13) | 0.29554 (12) | 0.23562 (11) | 0.0383 (2)          |
| C2   | 0.04121 (12) | 0.43215 (10) | 0.08574 (9)  | 0.03209 (18)        |
| C3   | −0.12671 (11)| 0.40482 (10) | 0.01398 (9)  | 0.03191 (18)        |
| C4   | 0.16818 (11) | 0.52677 (10) | 0.07290 (9)  | 0.03271 (18)        |
| C5   | 0.14913 (14) | 0.27582 (12) | 0.49964 (10) | 0.0399 (2)          |
| H5   | 0.1746       | 0.2134       | 0.5445       | 0.048*              |
| C6   | 0.08369 (13) | 0.35981 (11) | 0.16930 (10) | 0.0363 (2)          |
| C7   | 0.05679 (13) | −0.01820 (12)| 0.21266 (11) | 0.0395 (2)          |
| H7   | 0.1102       | −0.0643      | 0.1424       | 0.047*              |
| C8   | 0.41254 (13) | 0.20975 (14) | 0.35141 (12) | 0.0443 (2)          |
| H8   | 0.4682       | 0.3188       | 0.4105       | 0.053*              |
| C9   | −0.25754 (13)| 0.30944 (12) | 0.02455 (11) | 0.0403 (2)          |
| H9   | −0.2325      | 0.2646       | 0.0810       | 0.048*              |
| C10  | 0.33841 (13) | 0.55747 (14) | 0.14507 (12) | 0.0439 (2)          |
| H10  | 0.3673       | 0.5137       | 0.2019       | 0.053*              |
| C11  | −0.41805 (15)| 0.28305 (15) | −0.04652 (13)| 0.0495 (3)          |
| H11  | −0.5017      | 0.2200       | −0.0386      | 0.059*              |
| C12  | 0.45931 (14) | 0.64956 (16) | 0.13241 (14) | 0.0523 (3)          |
| H12  | 0.5698       | 0.6683       | 0.1805       | 0.063*              |
| C13  | −0.03051 (18)| 0.27784 (18) | 0.48561 (15) | 0.0568 (3)          |
| H13A | −0.1008      | 0.1751       | 0.4377       | 0.085*              |
| H13B | −0.0624      | 0.3302       | 0.4341       | 0.085*              |
| H13C | −0.0416      | 0.3295       | 0.5763       | 0.085*              |
| C14  | 0.45544 (17) | 0.16374 (19) | 0.21661 (16) | 0.0597 (3)          |
| H14A | 0.4160       | 0.2215       | 0.1731       | 0.090*              |
| H14B | 0.4037       | 0.0570       | 0.1552       | 0.090*              |
| H14C | 0.5735       | 0.1838       | 0.2369       | 0.090*              |
| C15  | 0.2658 (2)   | 0.43708 (16) | 0.59437 (14) | 0.0606 (3)          |
| H15A | 0.3788       | 0.4329       | 0.6096       | 0.091*              |
| H15B | 0.2446       | 0.4793       | 0.6817       | 0.091*              |
| H15C | 0.2471       | 0.5000       | 0.5511       | 0.091*              |
| C16  | 0.0579 (2)   | −0.10950 (15)| 0.29180 (16) | 0.0649 (4)          |
| H16A | 0.1703       | −0.1015      | 0.3362       | 0.097*              |
| H16B | −0.0002      | −0.2145      | 0.2282       | 0.097*              |
| H16C | 0.0042 | −0.0699 | 0.3608 | 0.097* |
| C17 | −0.12106 (17) | −0.03768 (17) | 0.13345 (16) | 0.0635 (4) |
| H17A | −0.1197 | 0.0169 | 0.0812 | 0.095* |
| H17B | −0.1803 | 0.0017 | 0.1981 | 0.095* |
| H17C | −0.1749 | −0.1439 | 0.0713 | 0.095* |
| C18 | 0.4854 (2) | 0.1290 (3) | 0.4289 (2) | 0.0851 (6) |
| H18A | 0.4509 | 0.1540 | 0.5098 | 0.128* |
| H18B | 0.6042 | 0.1615 | 0.4564 | 0.128* |
| H18C | 0.4469 | 0.0208 | 0.3692 | 0.128* |

*Atomic displacement parameters (Å²)*

| | $U^{11}$ | $U^{22}$ | $U^{33}$ | $U^{12}$ | $U^{13}$ | $U^{23}$ |
|---|---|---|---|---|---|---|
| Si1 | 0.03174 (14) | 0.03600 (15) | 0.03312 (14) | 0.01022 (9) | 0.00693 (9) | 0.02610 (12) |
| C1 | 0.0423 (5) | 0.0439 (5) | 0.0393 (5) | 0.0140 (4) | 0.0096 (4) | 0.0301 (4) |
| C2 | 0.0414 (4) | 0.0338 (4) | 0.0297 (4) | 0.0147 (3) | 0.0091 (3) | 0.0222 (4) |
| C3 | 0.0394 (4) | 0.0332 (4) | 0.0308 (4) | 0.0132 (3) | 0.0102 (3) | 0.0214 (4) |
| C4 | 0.0387 (4) | 0.0350 (4) | 0.0313 (4) | 0.0139 (3) | 0.0083 (3) | 0.0218 (4) |
| C5 | 0.0495 (5) | 0.0438 (5) | 0.0356 (4) | 0.0168 (4) | 0.0121 (4) | 0.0264 (4) |
| C6 | 0.0434 (5) | 0.0396 (5) | 0.0352 (4) | 0.0150 (4) | 0.0102 (4) | 0.0259 (4) |
| C7 | 0.0447 (5) | 0.0380 (5) | 0.0393 (5) | 0.0088 (4) | 0.0134 (4) | 0.0223 (4) |
| C8 | 0.0345 (4) | 0.0566 (6) | 0.0524 (6) | 0.0121 (4) | 0.0107 (4) | 0.0367 (5) |
| C9 | 0.0454 (5) | 0.0444 (5) | 0.0431 (5) | 0.0128 (4) | 0.0149 (4) | 0.0309 (4) |
| C10 | 0.0405 (5) | 0.0544 (6) | 0.0457 (5) | 0.0149 (4) | 0.0061 (4) | 0.0348 (5) |
| C11 | 0.0424 (5) | 0.0578 (7) | 0.0580 (7) | 0.0097 (5) | 0.0157 (5) | 0.0375 (6) |
| C12 | 0.0376 (5) | 0.0670 (8) | 0.0588 (7) | 0.0119 (5) | 0.0071 (5) | 0.0403 (6) |
| C13 | 0.0586 (7) | 0.0679 (8) | 0.0598 (7) | 0.0280 (6) | 0.0298 (6) | 0.0356 (7) |
| C14 | 0.0495 (6) | 0.0780 (9) | 0.0670 (8) | 0.0211 (6) | 0.0280 (6) | 0.0422 (7) |
| C15 | 0.0700 (8) | 0.0508 (7) | 0.0455 (6) | 0.0132 (6) | 0.0061 (6) | 0.0161 (6) |
| C16 | 0.0935 (11) | 0.0435 (6) | 0.0630 (8) | 0.0069 (6) | 0.0211 (7) | 0.0358 (6) |
| C17 | 0.0451 (6) | 0.0599 (8) | 0.0617 (8) | 0.0019 (5) | 0.0011 (5) | 0.0198 (6) |
| C18 | 0.0542 (8) | 0.1471 (18) | 0.1147 (14) | 0.0528 (10) | 0.0306 (9) | 0.1055 (15) |

*Geometric parameters (Å, °)*

| | | | |
|---|---|---|---|
| Si1—C1 | 1.8449 (11) | C10—H10 | 0.9300 |
| Si1—C8 | 1.8816 (11) | C11—C12[i] | 1.4120 (16) |
| Si1—C5 | 1.8843 (12) | C11—H11 | 0.9300 |
| Si1—C7 | 1.8877 (15) | C12—C11[i] | 1.4120 (16) |
| C1—C6 | 1.2031 (13) | C12—H12 | 0.9300 |
| C2—C4 | 1.4106 (14) | C13—H13A | 0.9600 |
| C2—C3 | 1.4114 (13) | C13—H13B | 0.9600 |
| C2—C6 | 1.4334 (12) | C13—H13C | 0.9600 |
| C3—C9 | 1.4242 (14) | C14—H14A | 0.9600 |
| C3—C4[i] | 1.4300 (12) | C14—H14B | 0.9600 |
| C4—C10 | 1.4233 (14) | C14—H14C | 0.9600 |
| C4—C3[i] | 1.4299 (11) | C15—H15A | 0.9600 |
| C5—C13 | 1.5229 (17) | C15—H15B | 0.9600 |
| C5—C15 | 1.533 (2) | C15—H15C | 0.9600 |
| C5—H5 | 0.9800 | C16—H16A | 0.9600 |
| C7—C16 | 1.5251 (16) | C16—H16B | 0.9600 |
| C7—C17 | 1.5278 (18) | C16—H16C | 0.9600 |
| C7—H7 | 0.9800 | C17—H17A | 0.9600 |
| C8—C18 | 1.5258 (17) | C17—H17B | 0.9600 |
| C8—C14 | 1.5285 (18) | C17—H17C | 0.9600 |
| C8—H8 | 0.9800 | C18—H18A | 0.9600 |
| C9—C11 | 1.3561 (16) | C18—H18B | 0.9600 |
| C9—H9 | 0.9300 | C18—H18C | 0.9600 |
| C10—C12 | 1.3580 (17) | | |
| C1—Si1—C8 | 106.22 (5) | C9—C11—H11 | 119.7 |
| C1—Si1—C5 | 107.65 (5) | C12[i]—C11—H11 | 119.7 |
| C8—Si1—C5 | 109.92 (6) | C10—C12—C11[i] | 120.32 (11) |
| C1—Si1—C7 | 107.78 (5) | C10—C12—H12 | 119.8 |
| C8—Si1—C7 | 112.32 (6) | C11[i]—C12—H12 | 119.8 |
| C5—Si1—C7 | 112.61 (6) | C5—C13—H13A | 109.5 |
| C6—C1—Si1 | 175.59 (10) | C5—C13—H13B | 109.5 |
| C4—C2—C3 | 120.57 (8) | H13A—C13—H13B | 109.5 |
| C4—C2—C6 | 119.59 (8) | C5—C13—H13C | 109.5 |
| C3—C2—C6 | 119.82 (9) | H13A—C13—H13C | 109.5 |
| C2—C3—C9 | 121.84 (8) | H13B—C13—H13C | 109.5 |
| C2—C3—C4[i] | 119.63 (8) | C8—C14—H14A | 109.5 |
| C9—C3—C4[i] | 118.53 (9) | C8—C14—H14B | 109.5 |
| C2—C4—C10 | 121.77 (8) | H14A—C14—H14B | 109.5 |
| C2—C4—C3[i] | 119.80 (8) | C8—C14—H14C | 109.5 |
| C10—C4—C3[i] | 118.43 (9) | H14A—C14—H14C | 109.5 |
| C13—C5—C15 | 110.31 (11) | H14B—C14—H14C | 109.5 |
| C13—C5—Si1 | 112.25 (9) | C5—C15—H15A | 109.5 |
| C15—C5—Si1 | 111.48 (9) | C5—C15—H15B | 109.5 |
| C13—C5—H5 | 107.5 | H15A—C15—H15B | 109.5 |
| C15—C5—H5 | 107.5 | C5—C15—H15C | 109.5 |
| Si1—C5—H5 | 107.5 | H15A—C15—H15C | 109.5 |
| C1—C6—C2 | 177.43 (11) | H15B—C15—H15C | 109.5 |
| C16—C7—C17 | 110.38 (11) | C7—C16—H16A | 109.5 |
| C16—C7—Si1 | 112.71 (8) | C7—C16—H16B | 109.5 |
| C17—C7—Si1 | 114.07 (9) | H16A—C16—H16B | 109.5 |
| C16—C7—H7 | 106.4 | C7—C16—H16C | 109.5 |
| C17—C7—H7 | 106.4 | H16A—C16—H16C | 109.5 |
| Si1—C7—H7 | 106.4 | H16B—C16—H16C | 109.5 |
| C18—C8—C14 | 110.76 (12) | C7—C17—H17A | 109.5 |
| C18—C8—Si1 | 112.81 (9) | C7—C17—H17B | 109.5 |
| C14—C8—Si1 | 113.35 (9) | H17A—C17—H17B | 109.5 |
| C18—C8—H8 | 106.5 | C7—C17—H17C | 109.5 |
| C14—C8—H8 | 106.5 | H17A—C17—H17C | 109.5 |
| Si1—C8—H8 | 106.5 | H17B—C17—H17C | 109.5 |
| C11—C9—C3 | 120.98 (9) | C8—C18—H18A | 109.5 |
| C11—C9—H9 | 119.5 | C8—C18—H18B | 109.5 |
| C3—C9—H9 | 119.5 | H18A—C18—H18B | 109.5 |
| C12—C10—C4 | 121.14 (10) | C8—C18—H18C | 109.5 |
| C12—C10—H10 | 119.4 | H18A—C18—H18C | 109.5 |
| C4—C10—H10 | 119.4 | H18B—C18—H18C | 109.5 |
| C9—C11—C12[i] | 120.59 (11) | | |
| | | | |
| C4—C2—C3—C9 | 179.79 (9) | C5—Si1—C7—C16 | −44.13 (11) |
| C6—C2—C3—C9 | 1.58 (14) | C1—Si1—C7—C17 | −35.82 (10) |
| C4—C2—C3—C4[i] | 0.26 (15) | C8—Si1—C7—C17 | −152.49 (9) |
| C6—C2—C3—C4[i] | −177.94 (8) | C5—Si1—C7—C17 | 82.76 (10) |
| C3—C2—C4—C10 | 179.40 (9) | C1—Si1—C8—C18 | 178.52 (12) |
| C6—C2—C4—C10 | −2.39 (15) | C5—Si1—C8—C18 | 62.33 (13) |
| C3—C2—C4—C3[i] | −0.26 (15) | C7—Si1—C8—C18 | −63.88 (13) |
| C6—C2—C4—C3[i] | 177.95 (8) | C1—Si1—C8—C14 | −54.59 (11) |
| C1—Si1—C5—C13 | 57.78 (10) | C5—Si1—C8—C14 | −170.77 (9) |
| C8—Si1—C5—C13 | 173.07 (8) | C7—Si1—C8—C14 | 63.01 (11) |
| C7—Si1—C5—C13 | −60.88 (10) | C2—C3—C9—C11 | −179.20 (10) |
| C1—Si1—C5—C15 | −66.56 (9) | C4[i]—C3—C9—C11 | 0.34 (16) |
| C8—Si1—C5—C15 | 48.73 (10) | C2—C4—C10—C12 | −179.58 (11) |
| C7—Si1—C5—C15 | 174.78 (8) | C3[i]—C4—C10—C12 | 0.08 (17) |
| C1—Si1—C7—C16 | −162.71 (10) | C3—C9—C11—C12[i] | −0.33 (19) |
| C8—Si1—C7—C16 | 80.62 (11) | C4—C10—C12—C11[i] | −0.1 (2) |

Symmetry code: (i) −x, −y+1, −z.

(a)

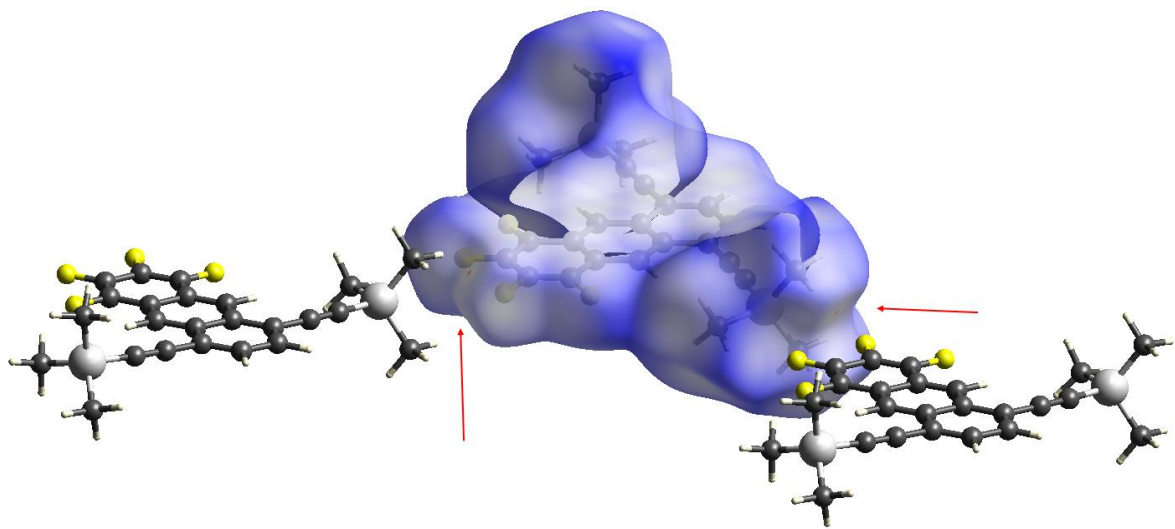

(b)

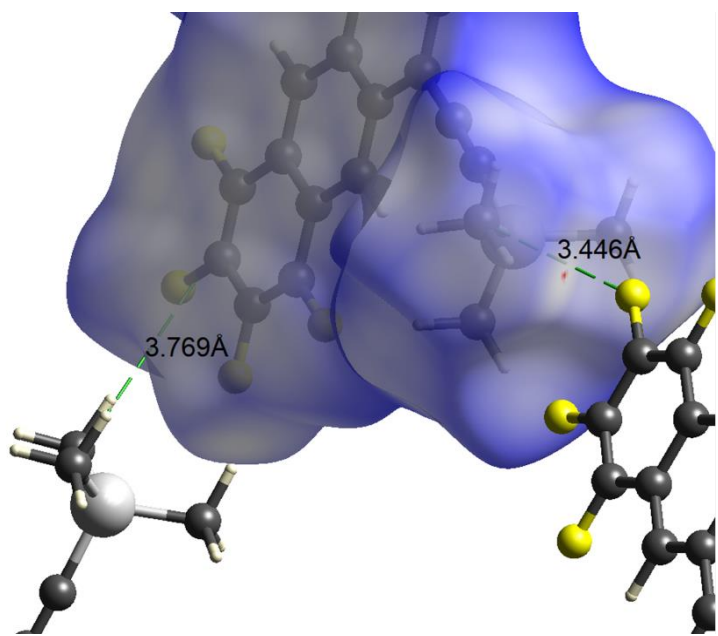

**Figure S1** F4 TMS ANT: (a) Hirshfeld surface mapped over $d_{norm}$ (surface transparency is enabled). The two neighbouring molecules, involved in $C_{aryl}-F \cdots H-C$ interactions, are also shown; red arrows indicate the presence of faint-red spots corresponding to $C_{aryl}-F \cdots H-C$ interactions; the red spots are more visible in the zoomed region shown in (b).

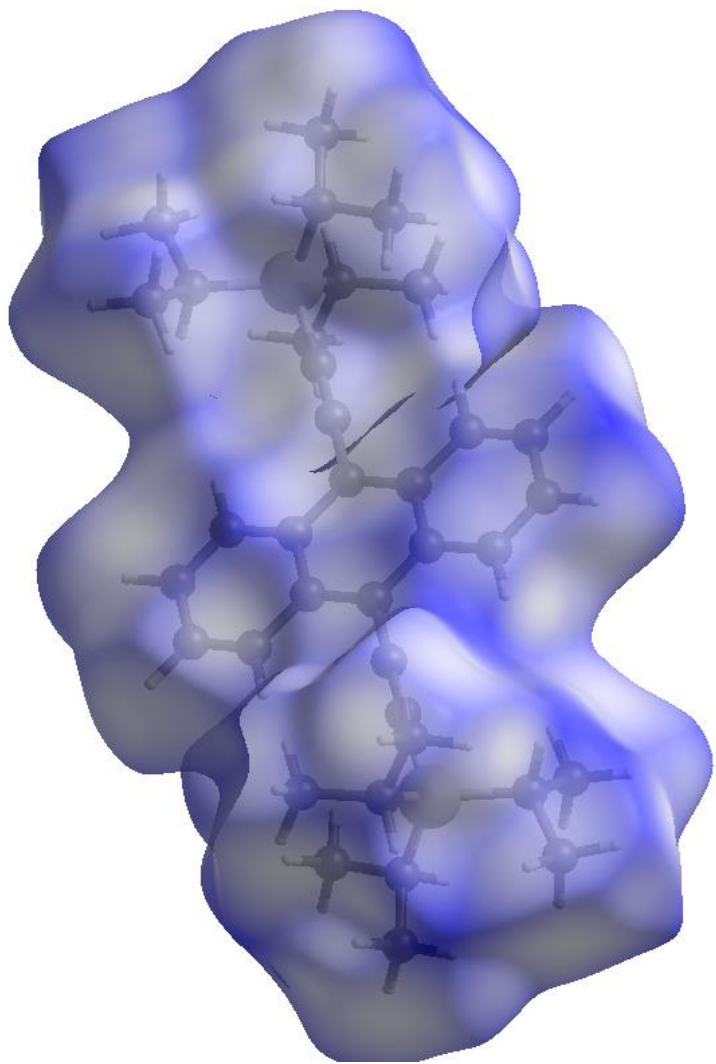

**Figure S2** TIPS ANT*p*: Hirshfeld surface mapped over $d_{norm}$ (surface transparency is enabled).

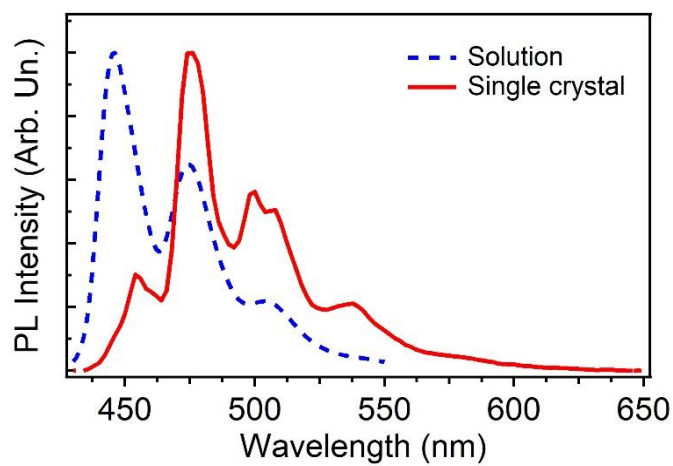

**Figure S3.** Solution (hexane, dashed line) and single crystals (continuous lines) emission spectra of TIPS ANT*p*.

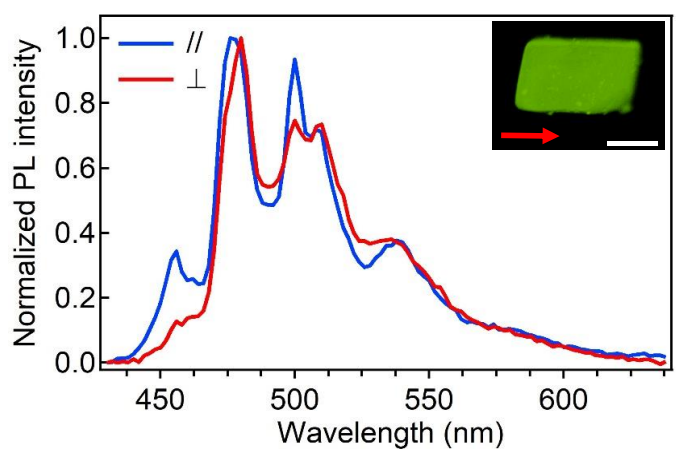

**Figure S4.** Polarized photoluminescence spectra of TIPS ANT*p* collected with the analyzer axis either parallel (blue continuous line) or perpendicular (red continuous line) to the short axis of the large face of the crystalline platelets. The intensities of the spectra are divided by their own maximum values. Inset: confocal fluorescence micrograph of a TIPS ANT*p* platelet. The direction of polarization of the excitation laser is highlighted by the red arrow. Scale bar: 50 μm.

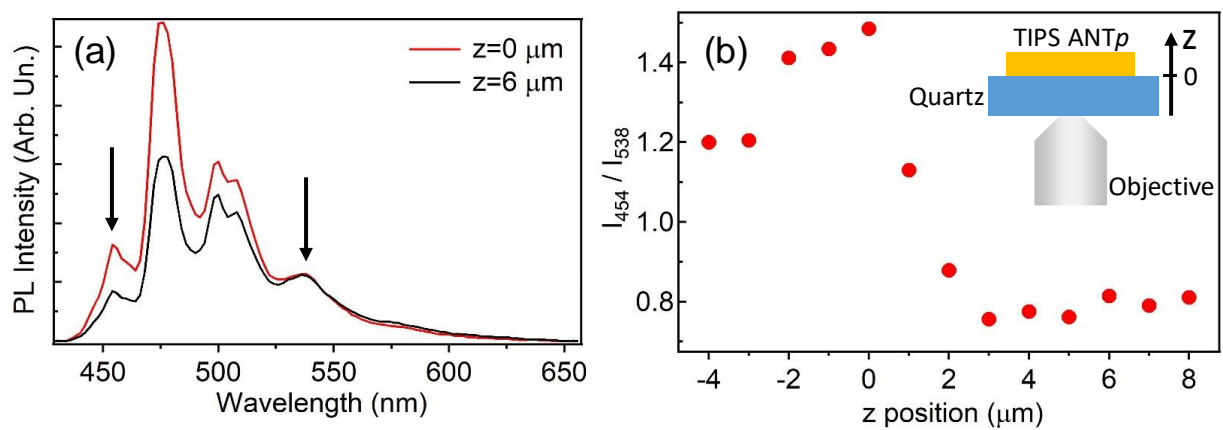

**Figure S5**. (a) Comparison of photoluminescence spectra of TIPS ANT*p*, measured by confocal microscopy at various objective positions (*i.e.*, positions along the vertical *z* axis, parallel to the crystal thickness). The spectra are normalized to the intensity of the peak at 538 nm. (b) *z*-dependence of the ratio between the PL intensity measured at $\lambda=454$ nm ($I_{454}$) and the one measured at $\lambda=538$ nm ($I_{538}$). These wavelengths are highlighted by vertical arrows in (a). Positive values of *z* correspond to positioning the excitation laser focal spot into the crystal thickness, whereas negative values correspond to the quartz substrate as depicted in the inset of (b).

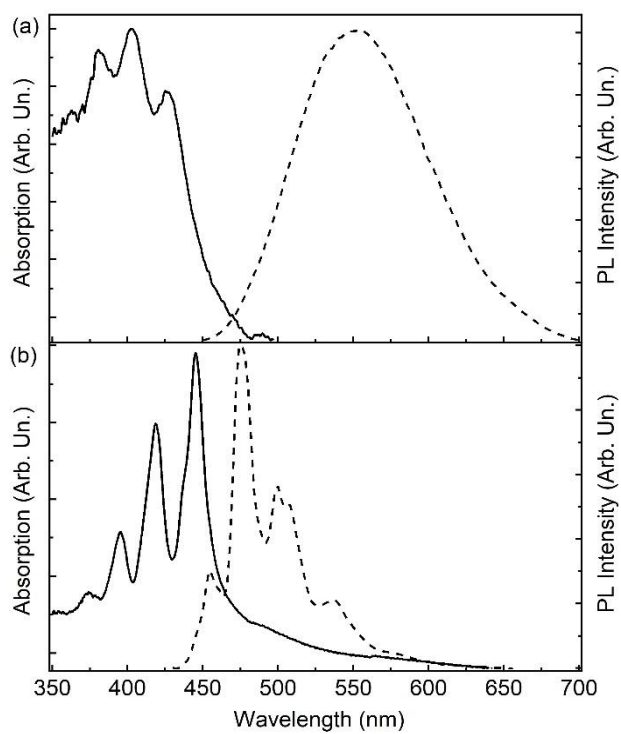

**Figure S6.** Absorption (continuous lines and left vertical scales) and photoluminescence (dashed lines and right vertical scales) spectra of single crystals of F4 TMS ANT (a) and TIPS ANT*p* (b).

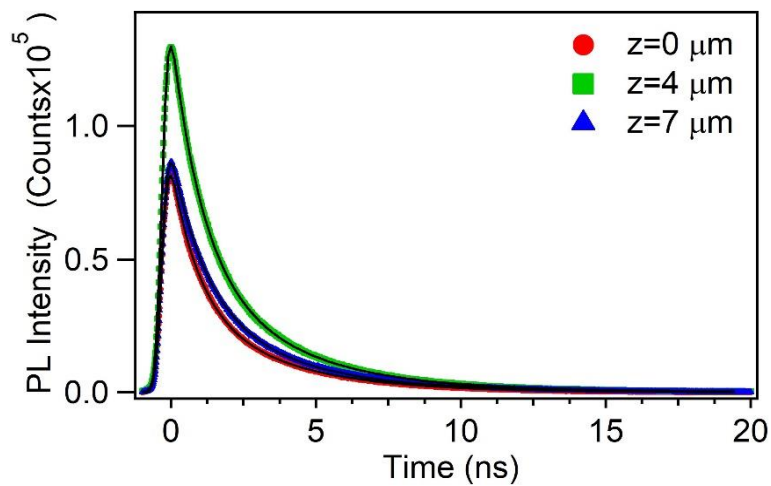

**Figure S7.** Comparison of photoluminescence decay of TIPS ANT*p* measured by confocal microscopy at various heights of the objective that focusses the excitation light and collects the emission. Positive values of *z* correspond to positioning the excitation laser focal spot into the crystal thickness as shown in Figure S5. The continuous black lines are fits to data by exponential functions convoluted with the instrumental response function. These measurements were performed by using the 100× objective with NA=1.3.